\documentclass[trackchanges,twocolumn]{aastex701}

\usepackage{amsmath}
\usepackage{amssymb}
\usepackage{pgfplots}
\usepackage{hyperref}
\usepackage{adjustbox}
\usepackage{tikz} 
\usepackage[normalem]{ulem}

\usetikzlibrary{positioning,calc}
\usepackage{pgfplots}
\usepackage{subcaption,graphicx}
\newsavebox{\tempfig}

\pgfplotsset{width=\paperwidth}
\pgfplotsset{compat=1.18}

\usepackage{xcolor} 

\newif\ifshowchanges
\showchangestrue  

\usepackage{environ} 

\hypersetup{
    colorlinks=true, 
    linkcolor=blue, 
    citecolor=blue, 
    urlcolor=blue 
}

\usepackage{ulem}

\begin{document}

\title{Neutron Economy and Freeze-out Dynamics in the Late-Time Cold r-Process}

\author[0000-0002-6373-7494]{Mengke Li}
\email{mengkel@berkeley.edu}
\affiliation{Department of Physics, University of California, Berkeley, CA 94720, USA}
\affiliation{Department of Physics and Astronomy, University of Notre Dame, Notre Dame, IN, 46656, USA}

\author[0000-0001-6307-9818]{Bradley S. Meyer}
\email{mbradle@clemson.edu}
\affiliation{
Department of Physics and Astronomy, Clemson University,
Clemson, SC 29634-0978, USA
}

\begin{abstract}
Standard r-process models often treat the freeze-out phase as a passive, uniform smoothing of the global abundance pattern.
We demonstrate, however, that the final pattern is instead actively sculpted by a dynamic ``neutron economy'' that operates well after the free neutron source is exhausted. 
By analyzing cold neutron star merger outflows, we identify a distinct neutron exporter-importer relationship: the second-peak region ($A \sim 130$) transitions early into a $\beta$-decay dominated regime, acting as a net exporter, while the third-peak ($A \sim 195$) and Rare Earth ($A \sim 164$) regions remain capture-dominated neutron importers. 
Targeted simulations reveal that the final peak structures emerge from a competition between two opposing forces: a global ``push'' driven by neutron capture shifting mass toward heavier nuclei, and a local ``pull'' from beta delayed neutron emission returning material back toward lighter masses. 
We identify $^{136}\text{Pd}$, $^{135}\text{Ag}$, and $^{130}\text{Ru}$ as the primary exporters driving this economy, while third-peak fine structure is governed by $N=126$ isotones (Sm to Tb).
Physically, this transport reflects a fundamental drive toward shell closure, where nuclei that overshoot the $N=82$ closed shell shed mass to migrate back toward this stable configuration, while nuclei in the third peak capture these redistributed neutrons to anchor themselves at the $N=126$ shell. 
Crucially, our models overproduce the odd-even staggering in the third peak compared to solar data. 
This discrepancy suggests current models likely overestimate the $\beta$-delayed neutron emission probabilities ($P_n$) for these $N=126$ nuclei, highlighting them as high-priority targets for theoretical refinement and future FRIB experiments.
\end{abstract}


\section{Introduction} 
The rapid neutron-capture process (r-process) has long been known to be responsible for synthesizing approximately half of the heavy elements in the universe \citep{1957RvMP...29..547B, 1965ApJS...11..121S}. 
Deciphering the origin of these elements is crucial for interpreting solar system abundances and constraining galactic chemical evolution.
Despite decades of study, however, robustly predicting r-process yields remains a challenge due to the coupled uncertainties of the astrophysical environment and the underlying nuclear physics. On the astrophysical front, proposed sites encompass a diverse range of extreme transients, most notably compact binary mergers and collapsars \citep{Abbott2017, Cote_2018, Miller_2020,Curtis_2023, Patel_2025}, which involve extreme multi-physics conditions that are computationally expensive to resolve.  Simultaneously, the r-process path traverses regions of the nuclear chart far from stability, where experimental constraints are scarce (e. g., \citealp{Horowitz_2019, Mumpower_2016}), forcing models to rely heavily on extrapolated nuclear properties.

Regardless of the specific site, the thermodynamic evolution of the ejecta imposes a characteristic sequence of nucleosynthetic regimes.
At early times, high temperatures and densities maintain the composition in nuclear statistical equilibrium (NSE) or quasi-statistical equilibrium (QSE). 
As the system expands and cools, it enters a phase where rapid neutron captures and photodissociations maintain $(n,\gamma)\rightleftharpoons(\gamma,n)$ equilibrium \citep{LI_2022}. Finally, as the temperature and neutron density drop further, this equilibrium breaks. 
It is during this critical ``freeze-out" transition, when neutron captures, photodissociations, and $\beta$-decays compete out of equilibrium, that the final abundance pattern is established (e. g., \citealp{Meyer_1994, Surman_1997,Arcones_2011, Eichler_2015, NISHIMURA_2016, Li_2025}). The specific mechanisms driving this late-time reshaping depend heavily on the neutron source. In extremely neutron-rich conditions (low $Y_e$), fission dictates the dynamics, creating a complex feedback loop of fragment cycling and prompt neutron emission that obscures other effects \citep{Eichler_2015, Vassh_2020}. In ``hot'' scenarios, sustained temperatures allow late-time photodissociation $(\gamma, n)$ to actively strip neutrons from material decaying back to stability. Regardless of these environment-specific sources, however, one late-time neutron source remains always-present: $\beta$-delayed neutron emission. 
Because the r-process frequently populates isotopes where the $\beta$-decay energy exceeds the neutron separation energy ($Q_\beta > S_n$), the subsequent decay cascade back toward stability triggers delayed neutron emission \citep{HALL_2021}.
Consequently, this channel acts as a universal neutron supplier active in every astrophysical scenario. Yet, in fissioning or hot environments, its precise shaping role is often difficult to disentangle from the massive neutron fluxes of other channels. 

In order to explore the role of beta-delayed neutron emission in the late-time evolution of r-process network dynamics, we consider the simplest astrophysical environment with the simplest freezeout neutron source: a ``cold'', moderately neutron-rich wind associated with neutron star mergers. In this regime, rapid cooling suppresses photodissociation, and the exclusion of extremely low-$Y_e$ material renders fission negligible. This leaves $\beta$-delayed neutron emission as the sole source of free neutrons during freeze-out. The released neutrons are not lost; they are efficiently recaptured by local species, reshaping the abundance distribution and shifting peak locations. 
Consequently, this scenario provides a clean laboratory for establishing the baseline physics of r-process pattern formation and determining the associated nuclear data sensitivities.

In this work, we frame this late-time phase through the lens of a dynamic ``neutron economy''. Rather than considering the free neutron density as a mere environmental background, we investigate the specific supply-and-demand relationships connecting the main r-process peaks. We show for the studied scenario (cold, moderate neutron-rich condition)
that the final abundance pattern is shaped by a distinct exporter-importer dynamic: the second-peak region ($A \approx 130$) transitions early into a $\beta$-decay dominated regime, acting as a net neutron exporter. Conversely, the Rare Earth and third-peak regions remain in a capture-dominated state, serving as net neutron importers. By quantifying this competition, we identify the specific nuclei that drive this economy and demonstrate how their interplay helps finalize the solar-like abundance structure.

\section{Methods}

\subsection{Simulation Framework}

Simulating $r$-process nucleosynthesis requires comprehensive nuclear data for over 7,000 exotic isotopes extending toward the neutron drip line. Because 
experimental measurements currently exist for only roughly 2,500 of these species, simulations must rely heavily on theoretical nuclear models for nuclei far from stability \citep{Erler2012}. 
As a result, nucleosynthesis predictions depend on both the underlying nuclear physics inputs (e.g., mass models, reaction rates, decay properties) and 
the assumed astrophysical trajectory. In this work, we adopt a representative 
baseline set of theoretical inputs to isolate the physical mechanics governing 
abundance peak migration and stabilization mechanism. While absolute 
quantitative features carry model-dependent uncertainties, systematically evaluating 
variations across different theoretical frameworks is beyond the scope of this study.

Nucleosynthesis calculations were performed using the \textsc{PRISM} (Portable Routines for Integrated nucleoSynthesis Modeling) reaction network \citep{Mumpower_2016}. \textsc{PRISM} is a single-zone network that solves the system of coupled differential equations governing the abundance evolution of all relevant nuclear species.

The baseline nuclear physics inputs are from the Finite Range Droplet Model (FRDM2012) \citep{FRDM2012}. Neutron-capture and neutron-induced fission rates are calculated using the Los Alamos statistical code CoH \citep{Kawano2016_CoH}, while $\beta$-decay half-lives and delayed neutron emission probabilities are adopted from the QRPA+HF framework of \citet{Mumpower_2016_bdne, Moller2003}. To maximize accuracy, we replace theoretical values with experimental data wherever available: experimental masses are from the Atomic Mass Evaluation (AME2020) \citep{Wang_2021_AME}, and decay properties are taken from the NUBASE2020 evaluation \citep{Kondev_2021_Nubase2020}. For fission fragment distributions, we assume a symmetric, two-particle split for all fissioning nuclei.

To investigate freeze-out dynamics across varying conditions, we employ a suite of four trajectories characteristic of neutron star merger ejecta:
(1) a high-entropy, low-$Y_e$ ($0.02$) outflow with vigorous fission \citep{just_2015};
(2) a high-entropy, moderate-$Y_e$ ($0.23$) outflow \cite{PRISM_wind};
(3) a cold, low-$Y_e$ ($0.13$) outflow with fission \citep{Lund_2024}; and
(4) a cold, moderate-$Y_e$ ($0.20$) outflow \citep{Lund_2024}.
While we examine abundance shifts in all scenarios, our detailed flow analysis focuses on the cold, non-fissioning case (Trajectory 4). This regime effectively isolates the impact of $\beta$-delayed neutron emission, disentangling it from the complicating feedback loops of fission cycling and late-time photodissociation.

\subsection{Analysis Methodology}
We quantify the availability of free neutrons relative to heavy seeds using the ratio $R \equiv Y_n / Y_h$, where $Y_n$ is the free neutron abundance and $Y_h$ is the total abundance of heavy nuclei ($Z>6, N>6$). We define the onset of the late-time freeze-out phase as the moment when this ratio drops below unity ($R < 1$). To characterize the reaction dynamics during this phase, we calculate effective timescales for neutron capture ($\tau_n$), photodissociation ($\tau_\gamma$), and $\beta$-decay ($\tau_\beta$) using abundance-weighted averages within specific mass regions. 

To strictly disentangle the specific drivers of the neutron economy in the most straightforward case, we performed a targeted suite of simulations based on the cold, non-fissioning trajectory. This suite isolates the contribution of specific mass regions to the global neutron flux:
\begin{enumerate}
    \item \textbf{Standard:} The baseline simulation using the full physical model with all nuclear inputs enabled.
    \item \textbf{Control:} A simulation identical to Standard except all $\beta$-delayed neutron emission is artificially suppressed.
    \item \textbf{2nd-Peak Test:} Delayed neutron emission is enabled \textit{only} for nuclei in the second-peak region ($120 < A < 150$) and suppressed elsewhere.
    \item \textbf{REP Test:} Delayed neutron emission is enabled \textit{only} for the Rare Earth Peak region ($150 \le A \leq 185$).
    \item \textbf{3rd-Peak Test:} Delayed neutron emission is enabled \textit{only} for the third-peak region ($186 \le A \le 200$). 
\end{enumerate}

Here, suppression of delayed neutron emission means that all $\beta$-decay channels that normally emit neutrons are directed into the standard, non-neutron-emitting $\beta^-$ channel ($P_0 = 1, P_n = 0$). Crucially, total $\beta$-decay half-lives ($\tau_{1/2}$) and decay rates remain constant and time-independent across the entire simulation; only the decay branching ratios are renormalized. 
By comparing these test cases against the Standard and Control runs, we can isolate the specific role of regional neutron export versus local import in shaping the final abundance pattern.


\section{Results} 

\subsection{The Onset of the Neutron Economy}
We define the onset of the ``neutron economy'' as the critical epoch where the free neutron-to-seed ratio drops to unity ($R \equiv Y_n/Y_h \approx 1$). Above this threshold, free neutrons are abundant, and each nucleus has many neutrons it can capture.  Below this threshold, the free neutrons become a scarce resource, and further abundance evolution is no longer driven by an external neutron flux but instead by redistribution within a closed system: neutrons are reallocated from some nuclei—through processes such as photodissociation, fission, or $\beta$-delayed emission—to others that effectively value those neutrons more, namely species with higher neutron binding energies and thus a greater capacity to retain them.

\begin{figure}[h!]
\centering
\includegraphics[width=\linewidth]{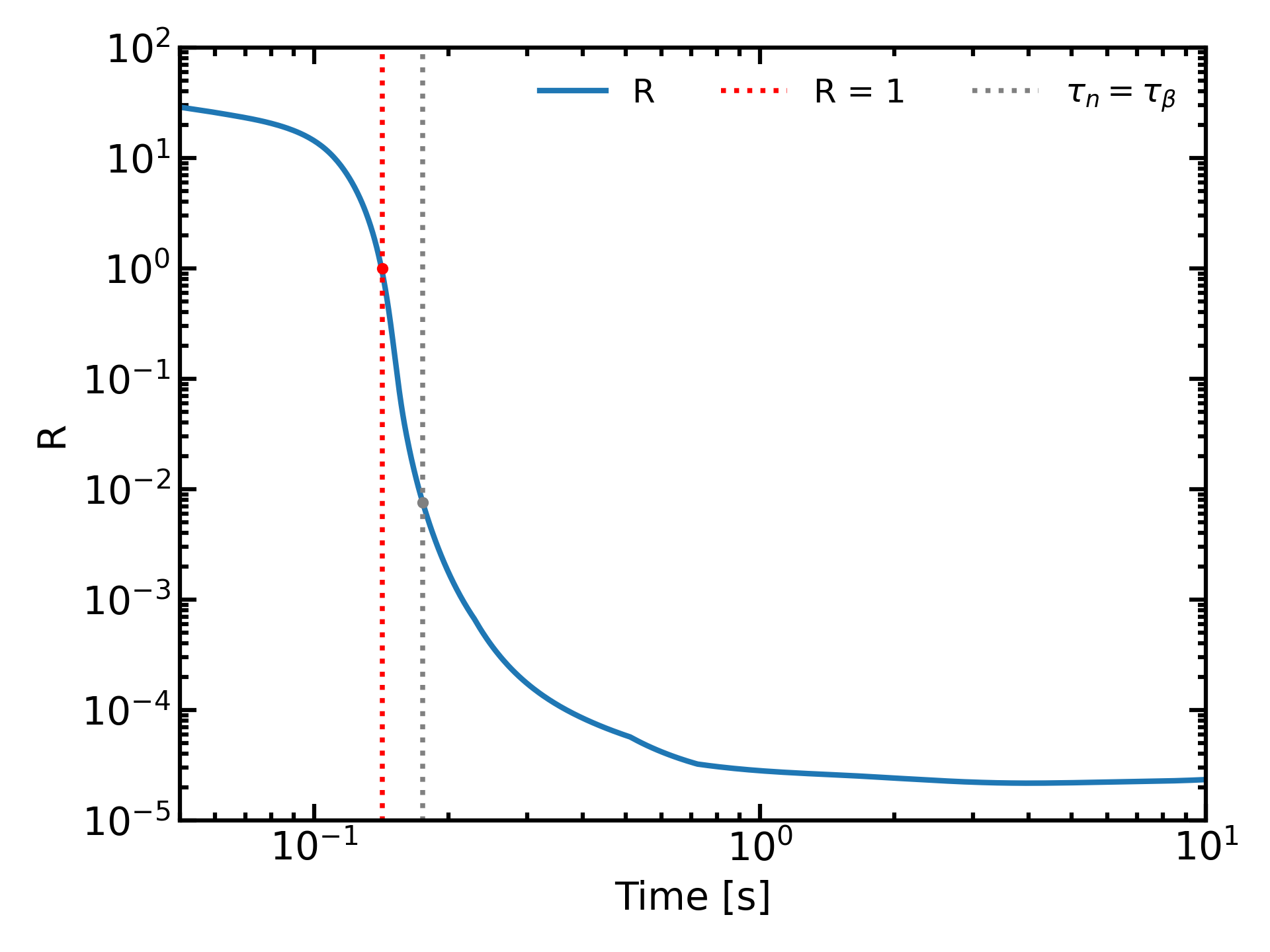}
\caption{
Time evolution of the neutron-to-seed ratio, defined as $R \equiv Y_n/Y_h$. 
The vertical dashed lines mark the boundaries of the neutron economy:
(1) the transition to the late-time phase where the ratio drops to unity ($R=1$), and 
(2) the freeze-out time where the global neutron capture timescale equals to the global beta-decay timescale ($\tau_n \approx \tau_\beta$).
}
\label{fig:R}   
\end{figure}

We track the neutron economy from the onset until the global $\tau_n = \tau_\beta$ timescale equivalence, at which point the final abundance pattern is largely established with only minor subsequent variations. 
To visualize the dynamics between these two critical time steps, Figure \ref{fig:R} illustrates the time evolution of $R$, which exhibits distinct behaviors across these regimes. 
Initially, $R$ declines gradually while the neutron abundance remains high. Once $R < 1$, however, the ratio decreases rapidly until the system reaches final freeze-out, when the neutron capture timescale exceeds the $\beta$-decay timescale ($\tau_n \gtrsim \tau_\beta$). 
Following this transition, the system becomes dominated by $\beta$-decay, and the abundance pattern stabilizes. 
Our analysis focuses on the dynamics within this specific window of rapid decline to determine how the final abundance pattern is shaped by this internal neutron exchange.

\begin{figure}[h!]\centering\includegraphics[width=\linewidth]{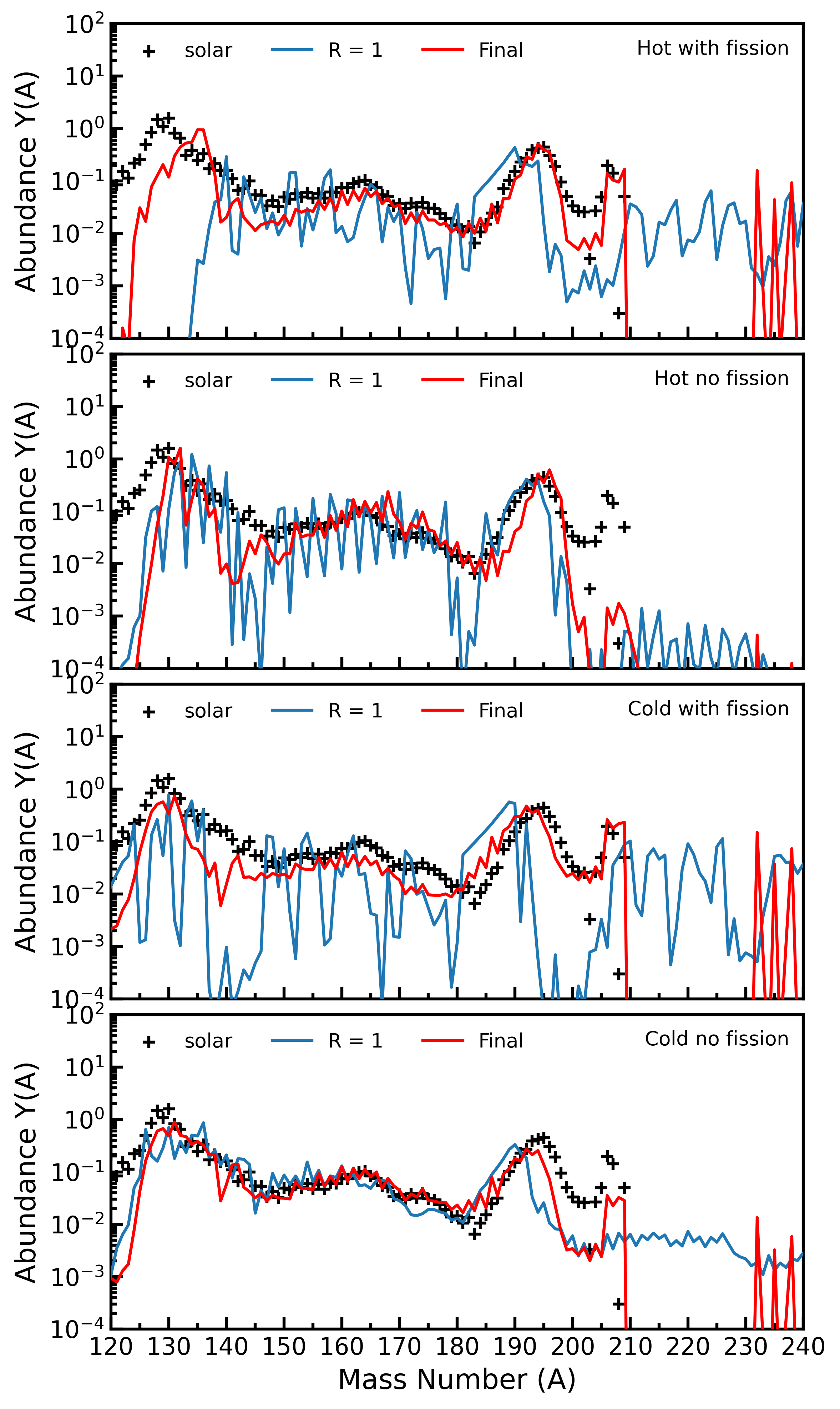}\caption{Evolution of abundance patterns from the onset of the neutron economy ($R=1$, blue) to final freeze-out (red). The third peak ($A \sim 195$) shifts toward heavier masses in all four scenarios \citep{just_2015, PRISM_wind, Lund_2024}. Panel (d) is selected for detailed analysis to isolate the role of $\beta$-delayed neutron emission.}\label{fig:ya_4condi}\end{figure}

Intuitively, the abundance pattern is expected to stabilize once the system drops below $R=1$, since it implies that there are insufficient free neutrons to support even a single capture per seed on average. However, Figure \ref{fig:ya_4condi} reveals a contradiction: significant mass transport persists well into this free-neutron-depleted phase. Most notably, the third r-process peak ($A \sim 195$) does not simply freeze in place; it migrates toward heavier isotopes, shifting by $2$ mass units between $R=1$ (blue) and final freeze-out (red).

This ``peak drift'' is robust across diverse astrophysical conditions, occurring in both hot, fission-recycling scenarios and cold, wind-like outflows (Figure \ref{fig:ya_4condi}) \citep{PRISM_wind, just_2015}. To understand this evolution in the most straightforward case, we focus our analysis on the cold, moderately neutron-rich trajectory (Panel d). In this ``cleaner'' environment, photodissociation and fission are suppressed, allowing us to isolate $\beta$-delayed neutron emission as the sole driver of this late-time reshaping.

\subsection{The Exporter-Importer Mechanism}

\begin{figure}[h!]\centering\includegraphics[width=\linewidth]{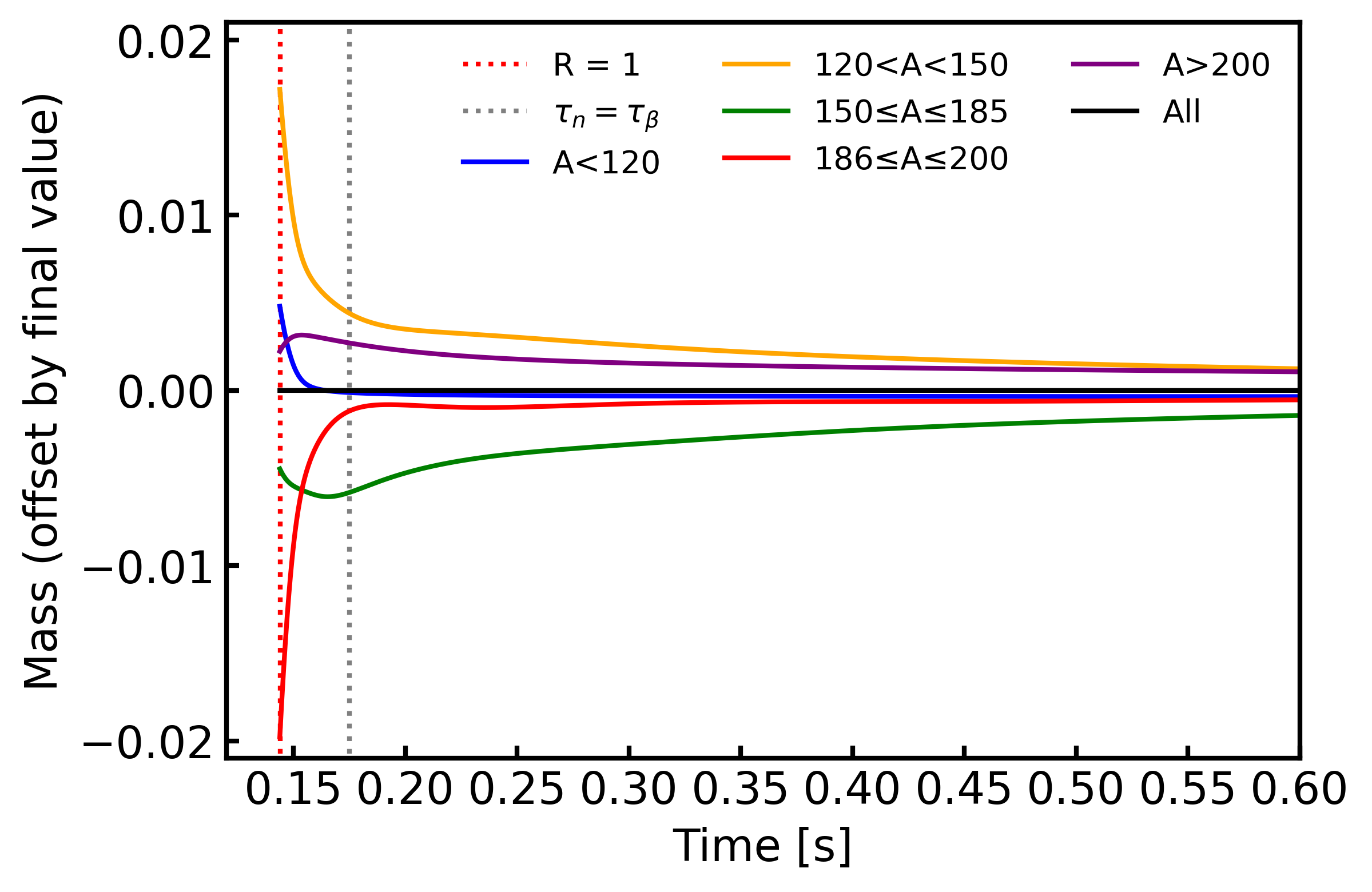}\caption{Time evolution of mass fractions across five regions in the cold, moderate $Y_e$ condition \citep{Lund_2024}. A clear mass transfer is evident after the onset of the neutron economy ($R=1$, dotted line): the Second Peak declines (acting as a supplier) while the Third Peak and REP grow (acting as consumers).}
\label{fig:mass_evol}
\end{figure}

To pinpoint the physical origin of the late-time peak drift, we track the evolution of the total mass fraction ($X = \sum A Y_A$) across five distinct mass regions \footnote{We define the regions as: Light ($A < 120$), Second Peak ($120 \leq A < 150$), Rare-Earth Peak/REP ($150 \leq A \leq 185$), Third Peak ($186 \leq A \leq 200$), and Trans-Lead ($A > 200$)}. Figure \ref{fig:mass_evol} reveals an inverse relationship between different mass regions immediately following the onset of the neutron economy ($R = 1$): as the total mass in the Second Peak region declines (solid orange line), the mass in the Third Peak region rises sharply (solid red line). Since fission is negligible in this scenario, this transfer implies a direct causality: the Second Peak acts as a net neutron supplier, shedding mass via $\beta$-delayed emission, while the Third Peak acts as a consumer, capturing these recycled neutrons to build up heavier isotopes.

To elucidate the mechanics of this transfer, we analyze the abundance-averaged reaction timescales,
\begin{equation}
    \tau_i = \frac{\sum_{(Z,A) \in B} Y(Z,A)}{\sum_{(Z,A) \in B} \lambda_i(Z,A) Y(Z,A)} \, ,
\end{equation}
where $Y(Z,A)$ is the abundance of isotope $(Z,A)$, $\lambda_i(Z,A)$ is the corresponding reaction rate for channel $i$, and $B$ denotes the specific mass bin under consideration. We evaluate these timescales for the primary competing channels: neutron capture ($\tau_n$) and $\beta$-decay ($\tau_\beta$).

\begin{figure*}[t]
\centering
\includegraphics[width=\linewidth]{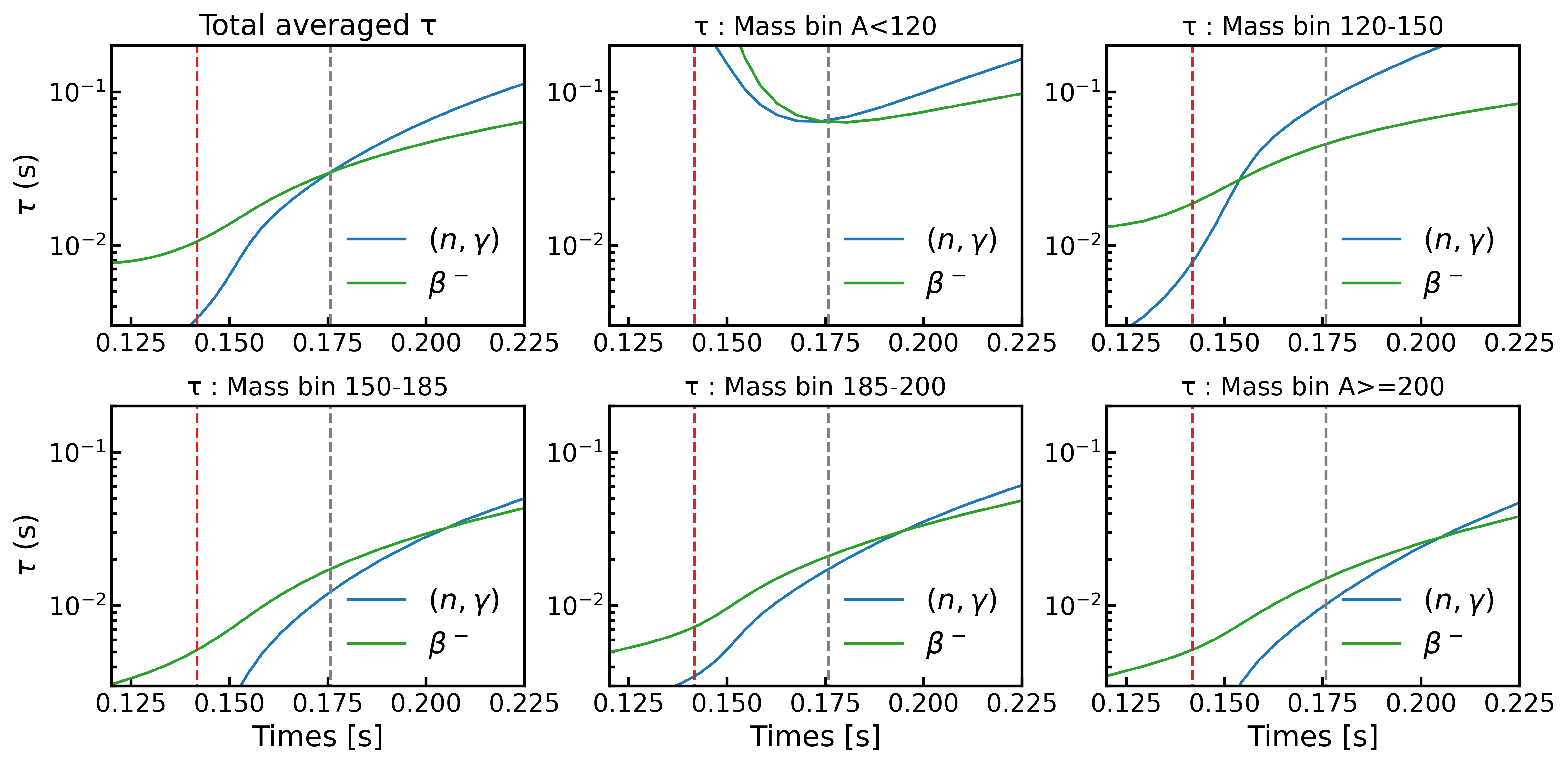}

\caption{Evolution of effective reaction timescales ($\tau_n$ vs. $\tau_\beta$). Photodisintegration timescales lie well beyond the relevant time window and are omitted for clarity.
The red vertical line stands for the timestep when $R = 1$; the gray dashed line marks the timestep $\tau_n = \tau_{\beta}$. While the global average (upper left) suggests a uniform capture environment, regional breakdowns reveal the mechanism: the Second Peak, shown in the upper right panel, becomes decay-dominated (exporter) while the Third Peak, shown in the bottom middle panel, remains capture-dominated (importer).}

\label{fig:tau_wdn}
\end{figure*}

The temporal evolution of these effective timescales during the late phase is presented in Figure \ref{fig:tau_wdn}. By comparing the relative dominance of $\tau_n$ and $\tau_\beta$ across different mass regions, we can map the dynamics of this neutron economy. The global average (Figure \ref{fig:tau_wdn}, upper-left panel) suggests that the system as a whole remains broadly in a capture-dominated regime ($\tau_n < \tau_\beta$) from the onset of the late-time phase ($R = 1$, red dashed line) until final freeze-out ($\tau_n = \tau_\beta$, gray dashed line). However, this global view masks a critical regional disparity. 

By decomposing the timescales by mass region (Figure \ref{fig:tau_wdn}, remaining panels), we identify the localized engines driving the neutron economy:

\begin{itemize}

\item \textbf{The Exporter (Second Peak):} This region transitions early into a decay-dominated regime ($\tau_\beta < \tau_n$). 
Here, nuclei actively decay back toward stability, releasing delayed neutrons that exit the region.

\item \textbf{The Importer (Third Peak):} This region remains firmly locked in a capture-dominated regime ($\tau_n \ll \tau_\beta$). 
The capture rates here are sufficiently fast to consume any free neutrons available, including those exported by the Second Peak.

\end{itemize}

This timescale mismatch creates a ``neutron gradient'', effectively pumping neutrons from the decaying Second Peak to the capturing Third Peak. This redistribution drives the systematic peak migration in opposite directions: the Second Peak is shifted toward lower masses, while the Third Peak is driven toward heavier isotopes.

\begin{figure}[h!]
\centering
\includegraphics[width=\linewidth]{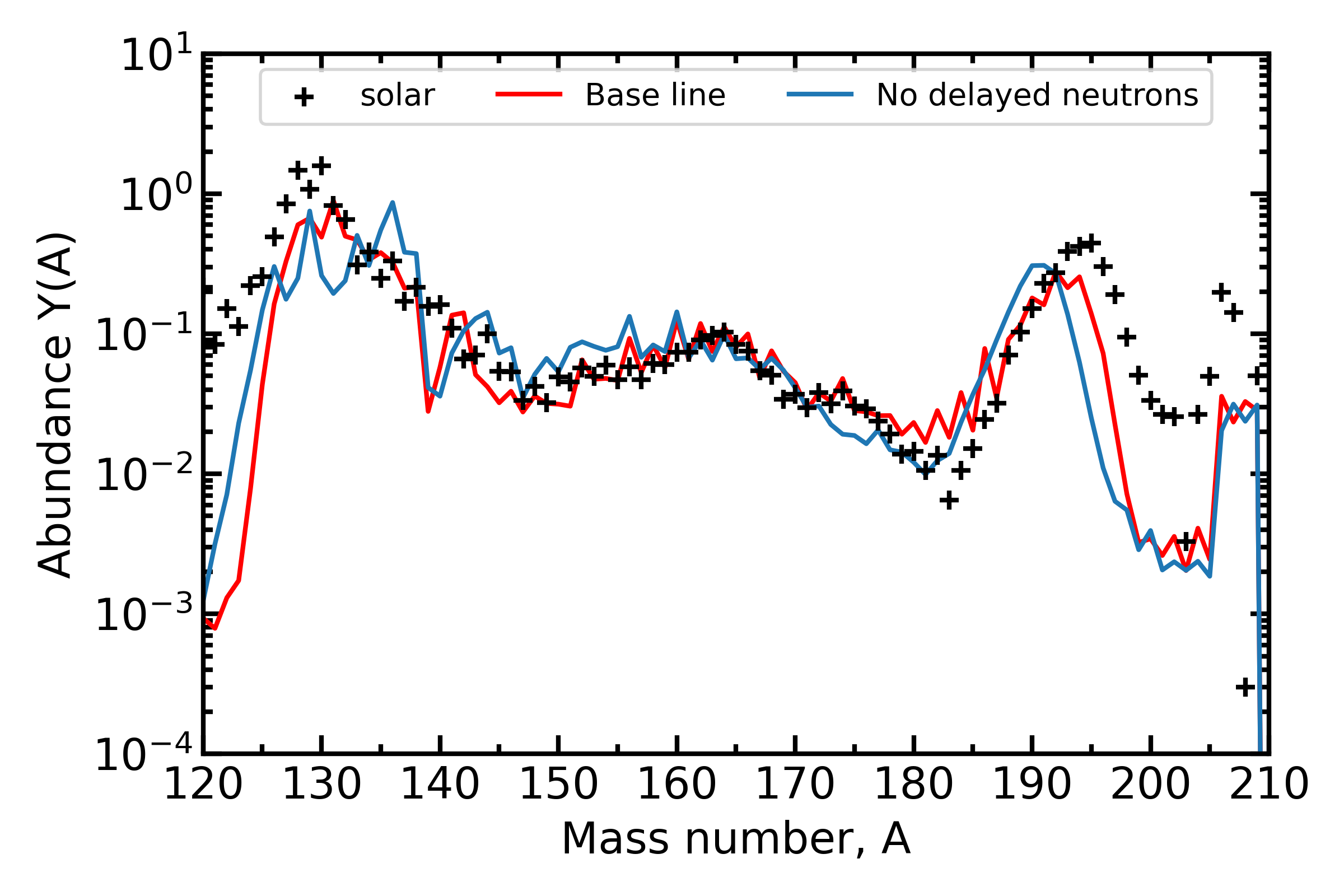}
\caption{
Comparison of final abundance patterns for the Standard (red) and Control (blue) simulations in the cold, moderate $Y_e$ condition \citep{Lund_2024}. The Control simulation has all $\beta$-delayed neutron emission artificially suppressed ($P_n=0$). The difference highlights the net mass transport driven specifically by delayed neutrons.
}
\label{fig:ya_w_wo}
\end{figure}

To confirm this mechanism, we compare our Standard model against a Control simulation where $\beta$-delayed neutron emission is modified such that all beta decay occurs through the no delayed neutron emission channel; that is, any channel with $P_n \ne 0$ has its probability moved to the no delayed neutron ($P_n=0$) channel. 
The simulated abundance results are shown in Figure \ref{fig:ya_w_wo}, the difference is profound. The Control simulation (blue) exhibits a flatter and wider Second peak and a Third peak that is shifted toward lower masses (left). 
By contrast, the Standard simulation (red) exhibits a narrowed Second Peak and a heavier Third Peak. This comparison confirms that the late-time reshaping is explicitly driven by the redistribution of mass via $\beta$-delayed neutron emission.

\begin{figure*}[t]
\centering
\includegraphics[width=\linewidth]{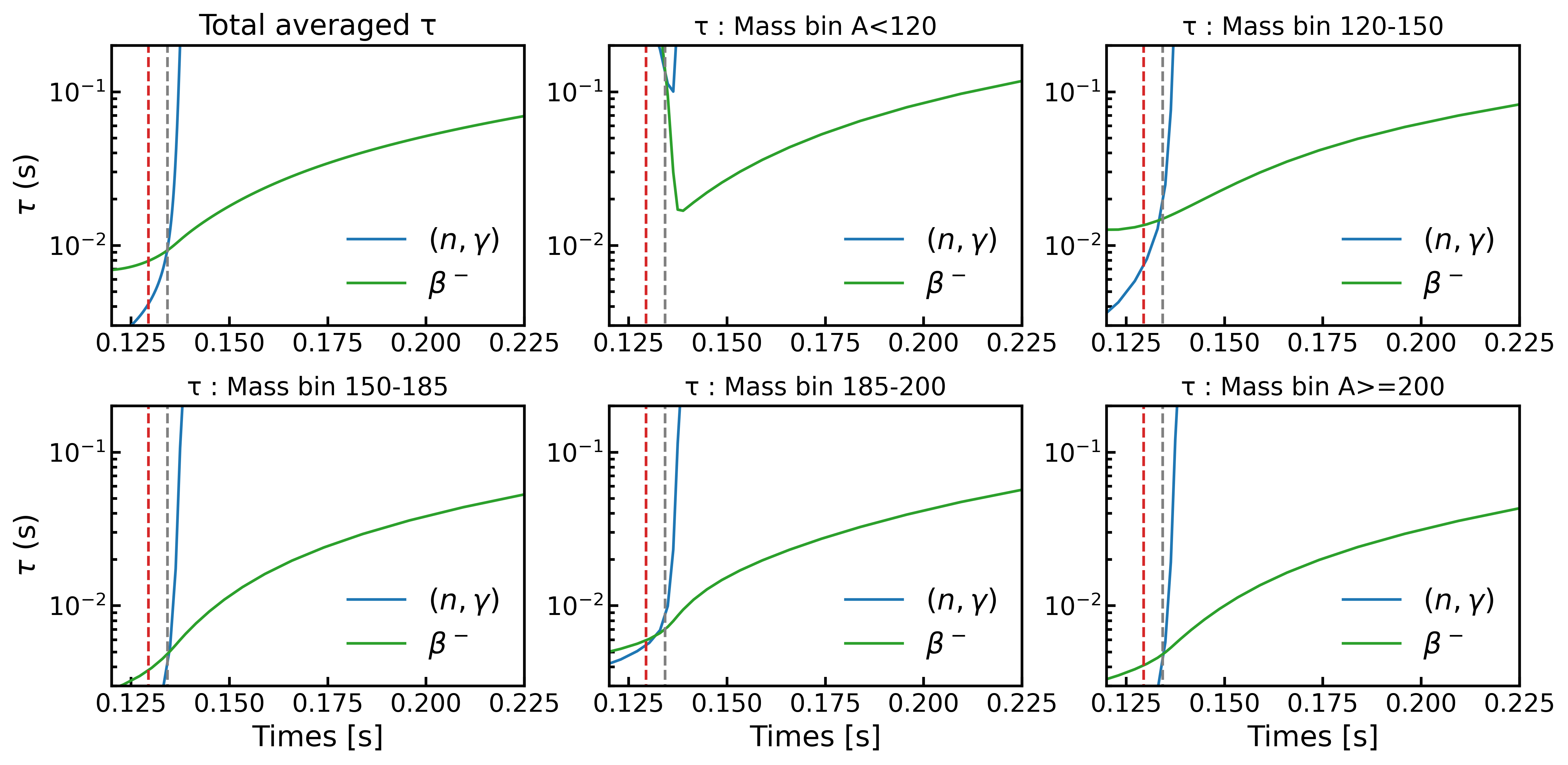}
\caption{
Same as Figure~\ref{fig:tau_wdn}, but for the control simulation with artificially suppressed $\beta$-delayed neutron emission ($P_n = 0$). The $x$-axis range is kept identical to Figure~\ref{fig:tau_wdn} to facilitate direct comparison.}
\label{fig:tau_ndn}
\end{figure*}

To understand why the redistribution mechanism fails in the Control case, we examine the reaction timescales in Figure \ref{fig:tau_ndn}, assessing whether the local imbalances between neutron capture and $\beta$-decay persist. First, the active time window shifts earlier and narrows significantly compared to Figure \ref{fig:tau_wdn}, reflecting the rapid depletion of free neutrons in the absence of late-time delayed emission. Then, as illustrated, the distinct regional behaviors vanish; instead, the crossing points where $\tau_n \approx \tau_\beta$ for individual mass regions align closely with the global trend (top left panel). This uniformity confirms that the exporter-importer dynamic has been extinguished. Without the injection of delayed neutrons to fuel local capture, the localized ``imbalance" between the second and third peaks disappears, and no late-time redistribution occurs. Consequently, the final abundance pattern lacks the essential reshaping required to match the solar r-process peaks.

\subsection{Push-pull mechanism}

To further analyze the shaping of the final abundance pattern, we performed a series of targeted simulations where $\beta$-delayed neutron emission was enabled exclusively for specific mass regions. 
These tests allow us to observe the distinct mechanical roles of each peak and define the forces at play.

\begin{figure}[h!]
\centering
\includegraphics[width=\linewidth]{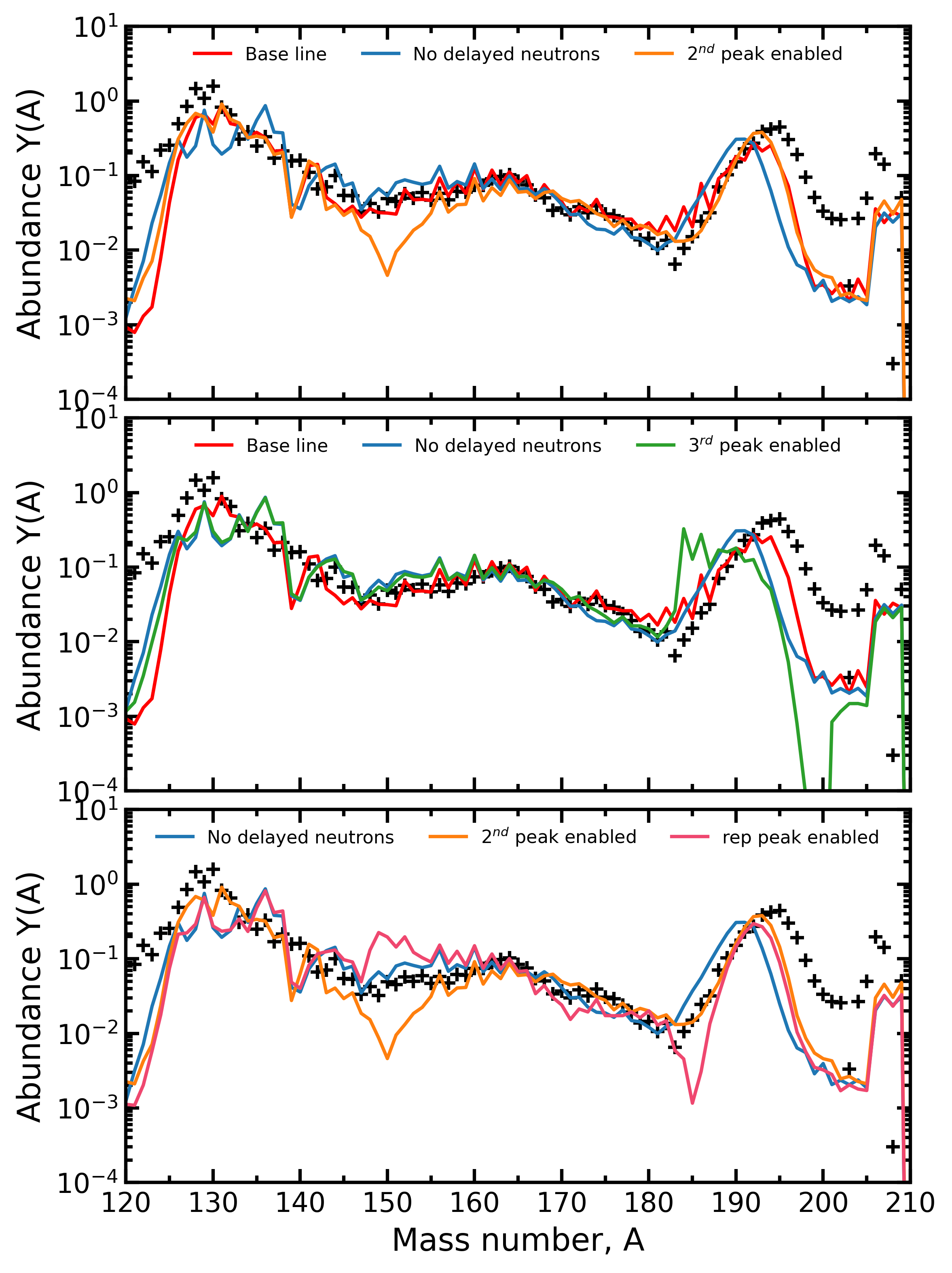}
\caption{
Deconstructing the Push-Pull mechanism. We compare the Standard simulation (red) and Control ($P_n=0$, blue) against targeted tests: The ``Exporter-Only" case (orange, top panel) isolates the global push to heavier masses. The ``Local-Only" case (green, middle panel) isolates the local pull and staggering. The REP test (pink) confirms this dynamic is universal (bottom panel).
}
\label{fig:control_simulation}
\end{figure}

First, we isolated the contribution of the exporter region by enabling delayed neutron emission only for the Second Peak. As shown in the top panel of Figure \ref{fig:control_simulation} (orange), this releases free neutrons that drive the capture path upward. 
The result is a transport of matter: the Second Peak sheds mass, and the Third Peak shifts significantly to the right (heavier masses). 
We identify this external, capture-driven force as the ``Global Push" since all nuclei can capture neutrons such that the abundances are ``pushed'' to higher mass.
However, the resulting Third Peak is not identical to the standard results, as it is too smooth and ``over-shifted" compared to the standard model, indicating that while the Global Push provides the bulk material, it cannot account for the final fine structure.

To find the missing shaping mechanism, we performed a test where delayed neutron emission was enabled only within the Third Peak region (middle panel, green). 
This isolation removes the masking effect of the external flux and reveals a different behavior.
Without imported neutrons from the Second Peak, the Third Peak cannot shift upward in mass; instead, its own internal delayed neutron emissions cause nuclei to decay back toward lighter masses. 
We identify this opposing force as the ``Local Pull'' since the abundances are pulled to lower mass but, owing to the low temperature and lack of fission in this scenario, the pull depends strictly on the local $\beta$-delayed emission properties.
Notably, this local decay carves out the characteristic odd-even staggering that was missing in the Global Push scenario. 
Crucially, this push-pull mechanism governs the entire mass distribution; in the Second Peak, the same competition occurs, but because the Local Pull significantly outweighs the Global Push, the region experiences a net drift toward lower masses. The final simulated abundance pattern is thus revealed to be the direct result of this system-wide competition.



Although the REP has a significantly lower overall abundance compared to the second and third peaks, its formation mechanism exhibits similarities. As shown in the bottom panel of Figure \ref{fig:control_simulation}, the formation of the REP is governed by the exact same dynamic. The ``Exporter-Only" neutron flux (orange) pushes the REP to higher masses, while local $\beta$ delayed neutron emission (pink) pulls it back to the left, lifting the low-mass shoulder. The final, solar-compatible position of the REP is achieved only when these opposing forces, the global push from the Second Peak and the local pull from internal beta decay, are both present.

\subsection{Decoupling Global Transport from Local Drivers}

\begin{figure*}[t]
\centering
\includegraphics[width=0.8\linewidth]{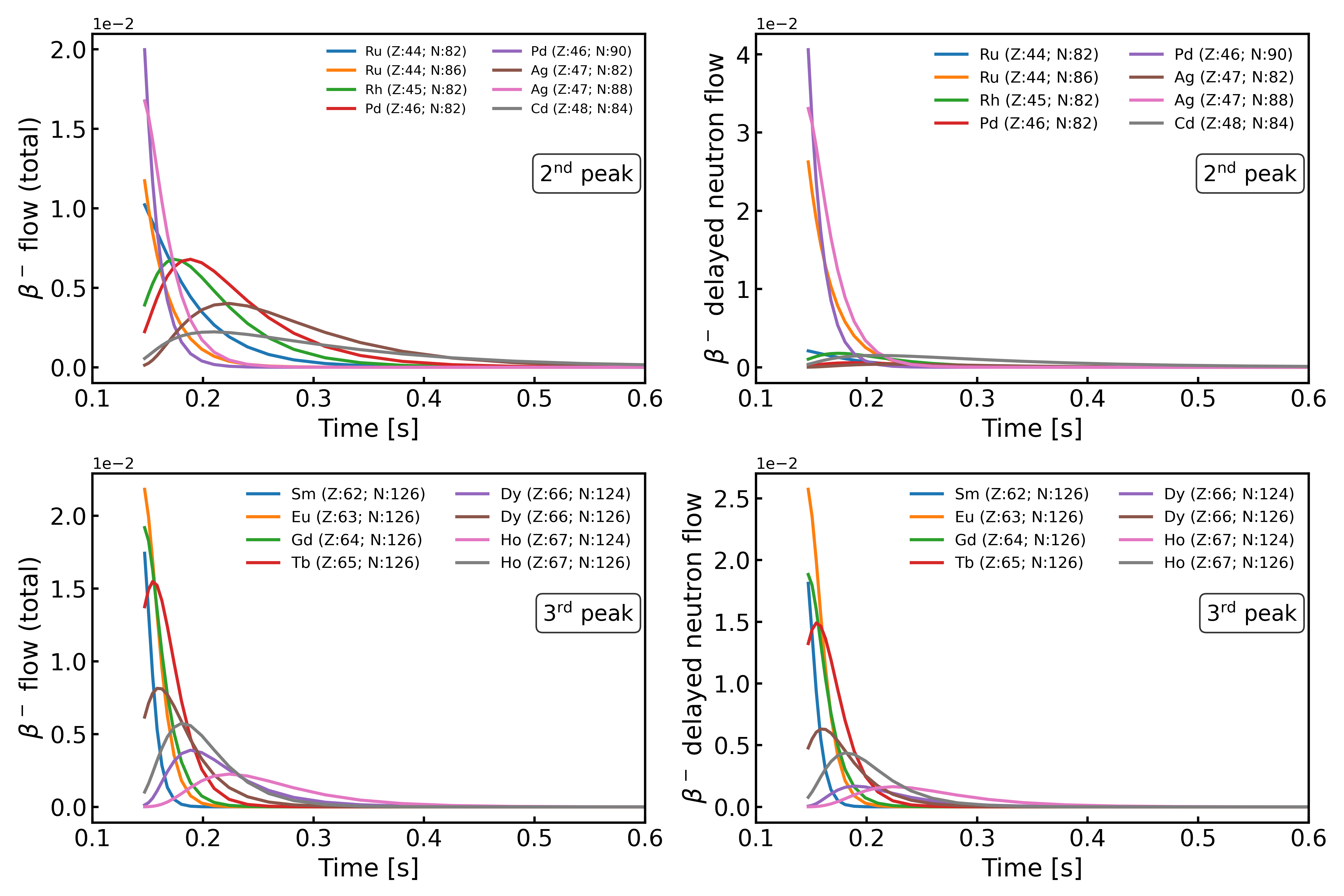}
\caption{Identification of key nuclei. We track the time evolution of the dominant contributors to the neutron flow. (Top) The "Exporters" in the Second Peak are identified as Pd, Ag, and Ru isotopes. 
(Bottom) The "Sculptors" of the Third Peak are identified as the $N=126$ isotones (Sm to Tb).}

\label{fig:flow}
\end{figure*}

Having established the mechanics of the neutron economy in the cold r-process under study, we now seek to identify the specific nuclei driving it. We focus our analysis on the Second and Third peak regions, as these two mass domains dominate the total r-process abundance and account for the vast majority of the active neutron exchange.
To pinpoint the key players, we calculate the effective delayed-neutron flow ($\lambda_\beta P_n Y$) for every isotope, isolating the species that combine high abundance, rapid decay, and high neutron emission probability ($P_n$). 
The top panel of Figure \ref{fig:flow} tracks the dominant contributors to the delayed neutron flux in the Second Peak region. 
A comparison between total decay activity (left panel) and effective neutron release (right) reveals that raw abundance is not the sole predictor of importance. 
For example, several abundant nuclei with high decay rates contribute little to the neutron economy due to negligible $P_n$ values, such as $^{126}$Ru and $^{127}$Rh. 
Filtering by effective flow, we identify the primary ``Exporters''—the engines of the global push—as $^{136}\text{Pd}$ ($Z=46$), $^{135}\text{Ag}$ ($Z=47$), and $^{130}\text{Ru}$ ($Z=44$). 
These specific isotopes effectively control the magnitude of the mass transfer to the heavier regions.

The bottom panel of Figure \ref{fig:flow} applies this analysis to the Third Peak to identify the sculptors responsible for the local shaping.
Here, we also distinguish between standard $\beta$-decay flow (left panel), which only moves material along an isobaric chain, while conserving mass number $A$, and $\beta$-delayed neutron flow (right panel), which directly modifies the mass distribution $Y(A)$ by shifting material to lower masses.
The analysis reveals that the dominant flows are driven exclusively by nuclei along the $N=126$ closed shell, specifically the isotones ranging from $^{188}\text{Sm}$ ($Z=62$) to $^{191}\text{Tb}$ ($Z=65$).
Notably, the top four nuclei with the 
largest $\beta$-decay flows are identical to those driving the largest 
$\beta$-delayed neutron flows.
Therefore, it is the specific decay chains of these neutron-rich isotones that generate the fine structure and characteristic odd-even staggering of the final peak.

This identification reveals that the ``push-pull'' dynamics are fundamentally driven by the system's tendency to return to closed shells. 
The ``Exporters" (Pd, Ag, Ru) are situated just beyond the $N=82$ shell closure ($N \approx 86-90$). 
As these nuclei decay rapidly back toward stability, their daughter isotopes tend to be less bound than $N=82$ nuclei a few mass units away.  This tends to favor a release of the ``excess'' neutrons that constitute the Global Push. 
Conversely, the Third Peak nuclei are formed at the $N=126$ shell closure. 
As they decay out of the closed shell, they utilize these imported neutrons to push back toward stability, with their own local emissions (the Local Pull) refining the final distribution.  The neutron economy is, in essence, a mechanism for transferring neutron excess from the post-$N=82$ region to the stabilizing $N=126$ region.

Finally, when we revisit the final abundance patterns (as seen in Figure \ref{fig:control_simulation}), a critical discrepancy emerges. The strong odd-even staggering generated by our Standard simulation is notably more pronounced than the smoother pattern observed in solar r-process residuals. Since our flow analysis has pinpointed the $N=126$ isotones ($^{188}\text{Sm}$--$^{191}\text{Tb}$) as the exclusive drivers of this local shaping, this excessive staggering suggests that current theoretical models likely overestimate the $\beta$-delayed neutron emission probabilities ($P_n$) for these specific nuclei.
This identifies the $N=126$ region as a high-priority target for future theoretical modeling and experimental campaigns.

\section{Conclusion and Discussion}

We have presented a framework for understanding the freeze-out phase of the r-process.
Previous study has seen the role of beta decay and the delayed neutron emission in reshaping the final abundance patterns \citep{NISHIMURA_2016, HALL_2021}, but the detailed dynamics are not thoroughly discussed.
In this work, by analyzing local reaction timescales and mass flows in a cold, moderate neutron rich condition, we reevaluate the standard assumption of a global freeze-out abundance smoothing. 
We demonstrated that the final abundance pattern is not merely a fossil of the early expansion, but is actively sculpted by a dynamic neutron economy that operates well after the external neutron source is exhausted. 

Our results show that the
shaping of the main r-process abundance peaks in the freeze-out stage is governed by a neutron exporter-importer relationship. 
Under the specific astrophysical conditions and baseline nuclear inputs adopted in this work, the Second Peak region ($A \sim 130$) transitions into a $\beta$-decay dominated regime early, acting as a net exporter of neutrons, while the Third Peak ($A \sim 195$) and Rare Earth regions remain capture-dominated neutron importers. 
This exchange manifests as a competition between a Global Push, which drives mass toward heavier isotopes, and a Local Pull from internal delayed neutron emission, which redistributes mass back toward lighter isotopes.

Physically, this competition is a direct consequence of nuclear structure, reflecting the universal drive to reach closed-shell configurations.
The neutron economy acts as the bridge allowing the unstable inventory beyond the $N=82$ shell (the exporters) to be transferred and locked into the stable $N=126$ configuration (the importers).

Crucially, our analysis uncovers discrepancies with observations across key mass regions. 
In the third peak, the $N=126$ ``sculptor'' nuclei ($^{188}\text{Sm}$--$^{191}\text{Tb}$) 
produce a stronger odd-even staggering than is observed in the solar abundance pattern. 
Additionally, the formation of the second peak exhibits high sensitivity to the decay data, resulting in a notable underproduction relative to solar data. 
These discrepancies highlight the urgent need for improved theoretical modeling and precision decay measurements far from stability.
While substantial experimental efforts have been achieved through measurement campaigns at RIBF \citep{Lorusso_2015, Phong_2022} and GSI/FAIR \citep{Caballero_2016}, continuous decay spectroscopy across international user facilities, such as FRIB 
\citep{Schatz_2022}, TRIUMF \citep{GARNSWORTHY_2019}, CERN-ISOLDE \citep{Borge_2017}, and HIAF \citep{Yang_2013}, remain essential to advancing our understanding of exotic nuclear structure and determining the production of $r$-process peaks.


Beyond shaping the final abundance pattern, this late-time neutron economy directly impacts the observable signatures of neutron star mergers \citep{Metzger_2010, Kasen_2017}. The specific production of heavy elements, particularly the lanthanides and actinides, determines how the expanding debris glows and cools over time \citep{Barnes_2013, Tanaka_2013}. Consequently, understanding these local decay flows is a critical ingredient for accurately modeling kilonova light curves. 
While the exact thermodynamic conditions of $r$-process ejecta remain subject to astrophysical uncertainties, the framework established here provides a foundation for delineating the late-time network dynamics that shapes the final r-process abundance pattern. Future studies will extend this framework to more complex environments, such as hot, fission-recycling scenarios, to provide a deeper understanding of and more precise predictions for the processes creating nature's heaviest elements.

\section{acknowledgements}
The authors thank the anonymous referee for carefully reading the manuscript and for providing insightful comments.
M. Li is supported by the NSF Grant PHY-2020275 (Network for Neutrinos, Nuclear Astrophysics and Symmetries (N3AS)). B. S. Meyer is supported by NASA grant 80NSSC20K0338.
We acknowledge the use of Gemini \citep{gemini_2025} for English language polishing. The authors reviewed, edited and take full responsibility for the final presented content.


\newpage
\bibliographystyle{aasjournalv7}
\bibliography{NE}

@article{Eichler_2015,
doi = {10.1088/0004-637X/808/1/30},
url = {https://doi.org/10.1088/0004-637X/808/1/30},
year = {2015},
month = {jul},
publisher = {The American Astronomical Society},
volume = {808},
number = {1},
pages = {30},
author = {Eichler, M. and Arcones, A. and Kelic, A. and Korobkin, O. and Langanke, K. and Marketin, T. and Martinez-Pinedo, G. and Panov, I. and Rauscher, T. and Rosswog, S. and Winteler, C. and Zinner, N. T. and Thielemann, F.-K.},
title = {THE ROLE OF FISSION IN NEUTRON STAR MERGERS AND ITS IMPACT ON THE r-PROCESS PEAKS},
journal = {The Astrophysical Journal}
}

@misc{gemini_2025,
      title={Gemini: A Family of Highly Capable Multimodal Models}, 
      author={Gemini Team and Rohan Anil and Sebastian Borgeaud and Jean-Baptiste Alayrac and Jiahui Yu and Radu Soricut and Johan Schalkwyk and Andrew M. Dai and Anja Hauth and Katie Millican and David Silver and Melvin Johnson and Ioannis Antonoglou and Julian Schrittwieser and Amelia Glaese and Jilin Chen and Emily Pitler and Timothy Lillicrap and Angeliki Lazaridou and Orhan Firat and James Molloy and Michael Isard and Paul R. Barham and Tom Hennigan and Benjamin Lee and Fabio Viola and Malcolm Reynolds and Yuanzhong Xu and Ryan Doherty and Eli Collins and Clemens Meyer and Eliza Rutherford and Erica Moreira and Kareem Ayoub and Megha Goel and Jack Krawczyk and Cosmo Du and Ed Chi and Heng-Tze Cheng and Eric Ni and Purvi Shah and Patrick Kane and Betty Chan and Manaal Faruqui and Aliaksei Severyn and Hanzhao Lin and YaGuang Li and Yong Cheng and Abe Ittycheriah and Mahdis Mahdieh and Mia Chen and Pei Sun and Dustin Tran and Sumit Bagri and Balaji Lakshminarayanan and Jeremiah Liu and Andras Orban and Fabian Güra and Hao Zhou and Xinying Song and Aurelien Boffy and Harish Ganapathy and Steven Zheng and HyunJeong Choe and Ágoston Weisz and Tao Zhu and Yifeng Lu and Siddharth Gopal and Jarrod Kahn and Maciej Kula and Jeff Pitman and Rushin Shah and Emanuel Taropa and Majd Al Merey and Martin Baeuml and Zhifeng Chen and Laurent El Shafey and Yujing Zhang and Olcan Sercinoglu and George Tucker and Enrique Piqueras and Maxim Krikun and Iain Barr and Nikolay Savinov and Ivo Danihelka and Becca Roelofs and Anaïs White and Anders Andreassen and Tamara von Glehn and Lakshman Yagati and Mehran Kazemi and Lucas Gonzalez and Misha Khalman and Jakub Sygnowski and Alexandre Frechette and Charlotte Smith and Laura Culp and Lev Proleev and Yi Luan and Xi Chen and James Lottes and Nathan Schucher and Federico Lebron and Alban Rrustemi and Natalie Clay and Phil Crone and Tomas Kocisky and Jeffrey Zhao and Bartek Perz and Dian Yu and Heidi Howard and Adam Bloniarz and Jack W. Rae and Han Lu and Laurent Sifre and Marcello Maggioni and Fred Alcober and Dan Garrette and Megan Barnes and Shantanu Thakoor and Jacob Austin and Gabriel Barth-Maron and William Wong and Rishabh Joshi and Rahma Chaabouni and Deeni Fatiha and Arun Ahuja and Gaurav Singh Tomar and Evan Senter and Martin Chadwick and Ilya Kornakov and Nithya Attaluri and Iñaki Iturrate and Ruibo Liu and Yunxuan Li and Sarah Cogan and Jeremy Chen and Chao Jia and Chenjie Gu and Qiao Zhang and Jordan Grimstad and Ale Jakse Hartman and Xavier Garcia and Thanumalayan Sankaranarayana Pillai and Jacob Devlin and Michael Laskin and Diego de Las Casas and Dasha Valter and Connie Tao and Lorenzo Blanco and Adrià Puigdomènech Badia and David Reitter and Mianna Chen and Jenny Brennan and Clara Rivera and Sergey Brin and Shariq Iqbal and Gabriela Surita and Jane Labanowski and Abhi Rao and Stephanie Winkler and Emilio Parisotto and Yiming Gu and Kate Olszewska and Ravi Addanki and Antoine Miech and Annie Louis and Denis Teplyashin and Geoff Brown and Elliot Catt and Jan Balaguer and Jackie Xiang and Pidong Wang and Zoe Ashwood and Anton Briukhov and Albert Webson and Sanjay Ganapathy and Smit Sanghavi and Ajay Kannan and Ming-Wei Chang and Axel Stjerngren and Josip Djolonga and Yuting Sun and Ankur Bapna and Matthew Aitchison and Pedram Pejman and Henryk Michalewski and Tianhe Yu and Cindy Wang and Juliette Love and Junwhan Ahn and Dawn Bloxwich and Kehang Han and Peter Humphreys and Thibault Sellam and James Bradbury and Varun Godbole and Sina Samangooei and Bogdan Damoc and Alex Kaskasoli and Sébastien M. R. Arnold and Vijay Vasudevan and Shubham Agrawal and Jason Riesa and Dmitry Lepikhin and Richard Tanburn and Srivatsan Srinivasan and Hyeontaek Lim and Sarah Hodkinson and Pranav Shyam and Johan Ferret and Steven Hand and Ankush Garg and Tom Le Paine and Jian Li and Yujia Li and Minh Giang and Alexander Neitz and Zaheer Abbas and Sarah York and Machel Reid and Elizabeth Cole and Aakanksha Chowdhery and Dipanjan Das and Dominika Rogozińska and Vitaliy Nikolaev and Pablo Sprechmann and Zachary Nado and Lukas Zilka and Flavien Prost and Luheng He and Marianne Monteiro and Gaurav Mishra and Chris Welty and Josh Newlan and Dawei Jia and Miltiadis Allamanis and Clara Huiyi Hu and Raoul de Liedekerke and Justin Gilmer and Carl Saroufim and Shruti Rijhwani and Shaobo Hou and Disha Shrivastava and Anirudh Baddepudi and Alex Goldin and Adnan Ozturel and Albin Cassirer and Yunhan Xu and Daniel Sohn and Devendra Sachan and Reinald Kim Amplayo and Craig Swanson and Dessie Petrova and Shashi Narayan and Arthur Guez and Siddhartha Brahma and Jessica Landon and Miteyan Patel and Ruizhe Zhao and Kevin Villela and Luyu Wang and Wenhao Jia and Matthew Rahtz and Mai Giménez and Legg Yeung and James Keeling and Petko Georgiev and Diana Mincu and Boxi Wu and Salem Haykal and Rachel Saputro and Kiran Vodrahalli and James Qin and Zeynep Cankara and Abhanshu Sharma and Nick Fernando and Will Hawkins and Behnam Neyshabur and Solomon Kim and Adrian Hutter and Priyanka Agrawal and Alex Castro-Ros and George van den Driessche and Tao Wang and Fan Yang and Shuo-yiin Chang and Paul Komarek and Ross McIlroy and Mario Lučić and Guodong Zhang and Wael Farhan and Michael Sharman and Paul Natsev and Paul Michel and Yamini Bansal and Siyuan Qiao and Kris Cao and Siamak Shakeri and Christina Butterfield and Justin Chung and Paul Kishan Rubenstein and Shivani Agrawal and Arthur Mensch and Kedar Soparkar and Karel Lenc and Timothy Chung and Aedan Pope and Loren Maggiore and Jackie Kay and Priya Jhakra and Shibo Wang and Joshua Maynez and Mary Phuong and Taylor Tobin and Andrea Tacchetti and Maja Trebacz and Kevin Robinson and Yash Katariya and Sebastian Riedel and Paige Bailey and Kefan Xiao and Nimesh Ghelani and Lora Aroyo and Ambrose Slone and Neil Houlsby and Xuehan Xiong and Zhen Yang and Elena Gribovskaya and Jonas Adler and Mateo Wirth and Lisa Lee and Music Li and Thais Kagohara and Jay Pavagadhi and Sophie Bridgers and Anna Bortsova and Sanjay Ghemawat and Zafarali Ahmed and Tianqi Liu and Richard Powell and Vijay Bolina and Mariko Iinuma and Polina Zablotskaia and James Besley and Da-Woon Chung and Timothy Dozat and Ramona Comanescu and Xiance Si and Jeremy Greer and Guolong Su and Martin Polacek and Raphaël Lopez Kaufman and Simon Tokumine and Hexiang Hu and Elena Buchatskaya and Yingjie Miao and Mohamed Elhawaty and Aditya Siddhant and Nenad Tomasev and Jinwei Xing and Christina Greer and Helen Miller and Shereen Ashraf and Aurko Roy and Zizhao Zhang and Ada Ma and Angelos Filos and Milos Besta and Rory Blevins and Ted Klimenko and Chih-Kuan Yeh and Soravit Changpinyo and Jiaqi Mu and Oscar Chang and Mantas Pajarskas and Carrie Muir and Vered Cohen and Charline Le Lan and Krishna Haridasan and Amit Marathe and Steven Hansen and Sholto Douglas and Rajkumar Samuel and Mingqiu Wang and Sophia Austin and Chang Lan and Jiepu Jiang and Justin Chiu and Jaime Alonso Lorenzo and Lars Lowe Sjösund and Sébastien Cevey and Zach Gleicher and Thi Avrahami and Anudhyan Boral and Hansa Srinivasan and Vittorio Selo and Rhys May and Konstantinos Aisopos and Léonard Hussenot and Livio Baldini Soares and Kate Baumli and Michael B. Chang and Adrià Recasens and Ben Caine and Alexander Pritzel and Filip Pavetic and Fabio Pardo and Anita Gergely and Justin Frye and Vinay Ramasesh and Dan Horgan and Kartikeya Badola and Nora Kassner and Subhrajit Roy and Ethan Dyer and Víctor Campos Campos and Alex Tomala and Yunhao Tang and Dalia El Badawy and Elspeth White and Basil Mustafa and Oran Lang and Abhishek Jindal and Sharad Vikram and Zhitao Gong and Sergi Caelles and Ross Hemsley and Gregory Thornton and Fangxiaoyu Feng and Wojciech Stokowiec and Ce Zheng and Phoebe Thacker and Çağlar Ünlü and Zhishuai Zhang and Mohammad Saleh and James Svensson and Max Bileschi and Piyush Patil and Ankesh Anand and Roman Ring and Katerina Tsihlas and Arpi Vezer and Marco Selvi and Toby Shevlane and Mikel Rodriguez and Tom Kwiatkowski and Samira Daruki and Keran Rong and Allan Dafoe and Nicholas FitzGerald and Keren Gu-Lemberg and Mina Khan and Lisa Anne Hendricks and Marie Pellat and Vladimir Feinberg and James Cobon-Kerr and Tara Sainath and Maribeth Rauh and Sayed Hadi Hashemi and Richard Ives and Yana Hasson and Eric Noland and Yuan Cao and Nathan Byrd and Le Hou and Qingze Wang and Thibault Sottiaux and Michela Paganini and Jean-Baptiste Lespiau and Alexandre Moufarek and Samer Hassan and Kaushik Shivakumar and Joost van Amersfoort and Amol Mandhane and Pratik Joshi and Anirudh Goyal and Matthew Tung and Andrew Brock and Hannah Sheahan and Vedant Misra and Cheng Li and Nemanja Rakićević and Mostafa Dehghani and Fangyu Liu and Sid Mittal and Junhyuk Oh and Seb Noury and Eren Sezener and Fantine Huot and Matthew Lamm and Nicola De Cao and Charlie Chen and Sidharth Mudgal and Romina Stella and Kevin Brooks and Gautam Vasudevan and Chenxi Liu and Mainak Chain and Nivedita Melinkeri and Aaron Cohen and Venus Wang and Kristie Seymore and Sergey Zubkov and Rahul Goel and Summer Yue and Sai Krishnakumaran and Brian Albert and Nate Hurley and Motoki Sano and Anhad Mohananey and Jonah Joughin and Egor Filonov and Tomasz Kępa and Yomna Eldawy and Jiawern Lim and Rahul Rishi and Shirin Badiezadegan and Taylor Bos and Jerry Chang and Sanil Jain and Sri Gayatri Sundara Padmanabhan and Subha Puttagunta and Kalpesh Krishna and Leslie Baker and Norbert Kalb and Vamsi Bedapudi and Adam Kurzrok and Shuntong Lei and Anthony Yu and Oren Litvin and Xiang Zhou and Zhichun Wu and Sam Sobell and Andrea Siciliano and Alan Papir and Robby Neale and Jonas Bragagnolo and Tej Toor and Tina Chen and Valentin Anklin and Feiran Wang and Richie Feng and Milad Gholami and Kevin Ling and Lijuan Liu and Jules Walter and Hamid Moghaddam and Arun Kishore and Jakub Adamek and Tyler Mercado and Jonathan Mallinson and Siddhinita Wandekar and Stephen Cagle and Eran Ofek and Guillermo Garrido and Clemens Lombriser and Maksim Mukha and Botu Sun and Hafeezul Rahman Mohammad and Josip Matak and Yadi Qian and Vikas Peswani and Pawel Janus and Quan Yuan and Leif Schelin and Oana David and Ankur Garg and Yifan He and Oleksii Duzhyi and Anton Älgmyr and Timothée Lottaz and Qi Li and Vikas Yadav and Luyao Xu and Alex Chinien and Rakesh Shivanna and Aleksandr Chuklin and Josie Li and Carrie Spadine and Travis Wolfe and Kareem Mohamed and Subhabrata Das and Zihang Dai and Kyle He and Daniel von Dincklage and Shyam Upadhyay and Akanksha Maurya and Luyan Chi and Sebastian Krause and Khalid Salama and Pam G Rabinovitch and Pavan Kumar Reddy M and Aarush Selvan and Mikhail Dektiarev and Golnaz Ghiasi and Erdem Guven and Himanshu Gupta and Boyi Liu and Deepak Sharma and Idan Heimlich Shtacher and Shachi Paul and Oscar Akerlund and François-Xavier Aubet and Terry Huang and Chen Zhu and Eric Zhu and Elico Teixeira and Matthew Fritze and Francesco Bertolini and Liana-Eleonora Marinescu and Martin Bölle and Dominik Paulus and Khyatti Gupta and Tejasi Latkar and Max Chang and Jason Sanders and Roopa Wilson and Xuewei Wu and Yi-Xuan Tan and Lam Nguyen Thiet and Tulsee Doshi and Sid Lall and Swaroop Mishra and Wanming Chen and Thang Luong and Seth Benjamin and Jasmine Lee and Ewa Andrejczuk and Dominik Rabiej and Vipul Ranjan and Krzysztof Styrc and Pengcheng Yin and Jon Simon and Malcolm Rose Harriott and Mudit Bansal and Alexei Robsky and Geoff Bacon and David Greene and Daniil Mirylenka and Chen Zhou and Obaid Sarvana and Abhimanyu Goyal and Samuel Andermatt and Patrick Siegler and Ben Horn and Assaf Israel and Francesco Pongetti and Chih-Wei "Louis" Chen and Marco Selvatici and Pedro Silva and Kathie Wang and Jackson Tolins and Kelvin Guu and Roey Yogev and Xiaochen Cai and Alessandro Agostini and Maulik Shah and Hung Nguyen and Noah Ó Donnaile and Sébastien Pereira and Linda Friso and Adam Stambler and Adam Kurzrok and Chenkai Kuang and Yan Romanikhin and Mark Geller and ZJ Yan and Kane Jang and Cheng-Chun Lee and Wojciech Fica and Eric Malmi and Qijun Tan and Dan Banica and Daniel Balle and Ryan Pham and Yanping Huang and Diana Avram and Hongzhi Shi and Jasjot Singh and Chris Hidey and Niharika Ahuja and Pranab Saxena and Dan Dooley and Srividya Pranavi Potharaju and Eileen O'Neill and Anand Gokulchandran and Ryan Foley and Kai Zhao and Mike Dusenberry and Yuan Liu and Pulkit Mehta and Ragha Kotikalapudi and Chalence Safranek-Shrader and Andrew Goodman and Joshua Kessinger and Eran Globen and Prateek Kolhar and Chris Gorgolewski and Ali Ibrahim and Yang Song and Ali Eichenbaum and Thomas Brovelli and Sahitya Potluri and Preethi Lahoti and Cip Baetu and Ali Ghorbani and Charles Chen and Andy Crawford and Shalini Pal and Mukund Sridhar and Petru Gurita and Asier Mujika and Igor Petrovski and Pierre-Louis Cedoz and Chenmei Li and Shiyuan Chen and Niccolò Dal Santo and Siddharth Goyal and Jitesh Punjabi and Karthik Kappaganthu and Chester Kwak and Pallavi LV and Sarmishta Velury and Himadri Choudhury and Jamie Hall and Premal Shah and Ricardo Figueira and Matt Thomas and Minjie Lu and Ting Zhou and Chintu Kumar and Thomas Jurdi and Sharat Chikkerur and Yenai Ma and Adams Yu and Soo Kwak and Victor Ähdel and Sujeevan Rajayogam and Travis Choma and Fei Liu and Aditya Barua and Colin Ji and Ji Ho Park and Vincent Hellendoorn and Alex Bailey and Taylan Bilal and Huanjie Zhou and Mehrdad Khatir and Charles Sutton and Wojciech Rzadkowski and Fiona Macintosh and Roopali Vij and Konstantin Shagin and Paul Medina and Chen Liang and Jinjing Zhou and Pararth Shah and Yingying Bi and Attila Dankovics and Shipra Banga and Sabine Lehmann and Marissa Bredesen and Zifan Lin and John Eric Hoffmann and Jonathan Lai and Raynald Chung and Kai Yang and Nihal Balani and Arthur Bražinskas and Andrei Sozanschi and Matthew Hayes and Héctor Fernández Alcalde and Peter Makarov and Will Chen and Antonio Stella and Liselotte Snijders and Michael Mandl and Ante Kärrman and Paweł Nowak and Xinyi Wu and Alex Dyck and Krishnan Vaidyanathan and Raghavender R and Jessica Mallet and Mitch Rudominer and Eric Johnston and Sushil Mittal and Akhil Udathu and Janara Christensen and Vishal Verma and Zach Irving and Andreas Santucci and Gamaleldin Elsayed and Elnaz Davoodi and Marin Georgiev and Ian Tenney and Nan Hua and Geoffrey Cideron and Edouard Leurent and Mahmoud Alnahlawi and Ionut Georgescu and Nan Wei and Ivy Zheng and Dylan Scandinaro and Heinrich Jiang and Jasper Snoek and Mukund Sundararajan and Xuezhi Wang and Zack Ontiveros and Itay Karo and Jeremy Cole and Vinu Rajashekhar and Lara Tumeh and Eyal Ben-David and Rishub Jain and Jonathan Uesato and Romina Datta and Oskar Bunyan and Shimu Wu and John Zhang and Piotr Stanczyk and Ye Zhang and David Steiner and Subhajit Naskar and Michael Azzam and Matthew Johnson and Adam Paszke and Chung-Cheng Chiu and Jaume Sanchez Elias and Afroz Mohiuddin and Faizan Muhammad and Jin Miao and Andrew Lee and Nino Vieillard and Jane Park and Jiageng Zhang and Jeff Stanway and Drew Garmon and Abhijit Karmarkar and Zhe Dong and Jong Lee and Aviral Kumar and Luowei Zhou and Jonathan Evens and William Isaac and Geoffrey Irving and Edward Loper and Michael Fink and Isha Arkatkar and Nanxin Chen and Izhak Shafran and Ivan Petrychenko and Zhe Chen and Johnson Jia and Anselm Levskaya and Zhenkai Zhu and Peter Grabowski and Yu Mao and Alberto Magni and Kaisheng Yao and Javier Snaider and Norman Casagrande and Evan Palmer and Paul Suganthan and Alfonso Castaño and Irene Giannoumis and Wooyeol Kim and Mikołaj Rybiński and Ashwin Sreevatsa and Jennifer Prendki and David Soergel and Adrian Goedeckemeyer and Willi Gierke and Mohsen Jafari and Meenu Gaba and Jeremy Wiesner and Diana Gage Wright and Yawen Wei and Harsha Vashisht and Yana Kulizhskaya and Jay Hoover and Maigo Le and Lu Li and Chimezie Iwuanyanwu and Lu Liu and Kevin Ramirez and Andrey Khorlin and Albert Cui and Tian LIN and Marcus Wu and Ricardo Aguilar and Keith Pallo and Abhishek Chakladar and Ginger Perng and Elena Allica Abellan and Mingyang Zhang and Ishita Dasgupta and Nate Kushman and Ivo Penchev and Alena Repina and Xihui Wu and Tom van der Weide and Priya Ponnapalli and Caroline Kaplan and Jiri Simsa and Shuangfeng Li and Olivier Dousse and Fan Yang and Jeff Piper and Nathan Ie and Rama Pasumarthi and Nathan Lintz and Anitha Vijayakumar and Daniel Andor and Pedro Valenzuela and Minnie Lui and Cosmin Paduraru and Daiyi Peng and Katherine Lee and Shuyuan Zhang and Somer Greene and Duc Dung Nguyen and Paula Kurylowicz and Cassidy Hardin and Lucas Dixon and Lili Janzer and Kiam Choo and Ziqiang Feng and Biao Zhang and Achintya Singhal and Dayou Du and Dan McKinnon and Natasha Antropova and Tolga Bolukbasi and Orgad Keller and David Reid and Daniel Finchelstein and Maria Abi Raad and Remi Crocker and Peter Hawkins and Robert Dadashi and Colin Gaffney and Ken Franko and Anna Bulanova and Rémi Leblond and Shirley Chung and Harry Askham and Luis C. Cobo and Kelvin Xu and Felix Fischer and Jun Xu and Christina Sorokin and Chris Alberti and Chu-Cheng Lin and Colin Evans and Alek Dimitriev and Hannah Forbes and Dylan Banarse and Zora Tung and Mark Omernick and Colton Bishop and Rachel Sterneck and Rohan Jain and Jiawei Xia and Ehsan Amid and Francesco Piccinno and Xingyu Wang and Praseem Banzal and Daniel J. Mankowitz and Alex Polozov and Victoria Krakovna and Sasha Brown and MohammadHossein Bateni and Dennis Duan and Vlad Firoiu and Meghana Thotakuri and Tom Natan and Matthieu Geist and Ser tan Girgin and Hui Li and Jiayu Ye and Ofir Roval and Reiko Tojo and Michael Kwong and James Lee-Thorp and Christopher Yew and Danila Sinopalnikov and Sabela Ramos and John Mellor and Abhishek Sharma and Kathy Wu and David Miller and Nicolas Sonnerat and Denis Vnukov and Rory Greig and Jennifer Beattie and Emily Caveness and Libin Bai and Julian Eisenschlos and Alex Korchemniy and Tomy Tsai and Mimi Jasarevic and Weize Kong and Phuong Dao and Zeyu Zheng and Frederick Liu and Fan Yang and Rui Zhu and Tian Huey Teh and Jason Sanmiya and Evgeny Gladchenko and Nejc Trdin and Daniel Toyama and Evan Rosen and Sasan Tavakkol and Linting Xue and Chen Elkind and Oliver Woodman and John Carpenter and George Papamakarios and Rupert Kemp and Sushant Kafle and Tanya Grunina and Rishika Sinha and Alice Talbert and Diane Wu and Denese Owusu-Afriyie and Cosmo Du and Chloe Thornton and Jordi Pont-Tuset and Pradyumna Narayana and Jing Li and Saaber Fatehi and John Wieting and Omar Ajmeri and Benigno Uria and Yeongil Ko and Laura Knight and Amélie Héliou and Ning Niu and Shane Gu and Chenxi Pang and Yeqing Li and Nir Levine and Ariel Stolovich and Rebeca Santamaria-Fernandez and Sonam Goenka and Wenny Yustalim and Robin Strudel and Ali Elqursh and Charlie Deck and Hyo Lee and Zonglin Li and Kyle Levin and Raphael Hoffmann and Dan Holtmann-Rice and Olivier Bachem and Sho Arora and Christy Koh and Soheil Hassas Yeganeh and Siim Põder and Mukarram Tariq and Yanhua Sun and Lucian Ionita and Mojtaba Seyedhosseini and Pouya Tafti and Zhiyu Liu and Anmol Gulati and Jasmine Liu and Xinyu Ye and Bart Chrzaszcz and Lily Wang and Nikhil Sethi and Tianrun Li and Ben Brown and Shreya Singh and Wei Fan and Aaron Parisi and Joe Stanton and Vinod Koverkathu and Christopher A. Choquette-Choo and Yunjie Li and TJ Lu and Abe Ittycheriah and Prakash Shroff and Mani Varadarajan and Sanaz Bahargam and Rob Willoughby and David Gaddy and Guillaume Desjardins and Marco Cornero and Brona Robenek and Bhavishya Mittal and Ben Albrecht and Ashish Shenoy and Fedor Moiseev and Henrik Jacobsson and Alireza Ghaffarkhah and Morgane Rivière and Alanna Walton and Clément Crepy and Alicia Parrish and Zongwei Zhou and Clement Farabet and Carey Radebaugh and Praveen Srinivasan and Claudia van der Salm and Andreas Fidjeland and Salvatore Scellato and Eri Latorre-Chimoto and Hanna Klimczak-Plucińska and David Bridson and Dario de Cesare and Tom Hudson and Piermaria Mendolicchio and Lexi Walker and Alex Morris and Matthew Mauger and Alexey Guseynov and Alison Reid and Seth Odoom and Lucia Loher and Victor Cotruta and Madhavi Yenugula and Dominik Grewe and Anastasia Petrushkina and Tom Duerig and Antonio Sanchez and Steve Yadlowsky and Amy Shen and Amir Globerson and Lynette Webb and Sahil Dua and Dong Li and Surya Bhupatiraju and Dan Hurt and Haroon Qureshi and Ananth Agarwal and Tomer Shani and Matan Eyal and Anuj Khare and Shreyas Rammohan Belle and Lei Wang and Chetan Tekur and Mihir Sanjay Kale and Jinliang Wei and Ruoxin Sang and Brennan Saeta and Tyler Liechty and Yi Sun and Yao Zhao and Stephan Lee and Pandu Nayak and Doug Fritz and Manish Reddy Vuyyuru and John Aslanides and Nidhi Vyas and Martin Wicke and Xiao Ma and Evgenii Eltyshev and Nina Martin and Hardie Cate and James Manyika and Keyvan Amiri and Yelin Kim and Xi Xiong and Kai Kang and Florian Luisier and Nilesh Tripuraneni and David Madras and Mandy Guo and Austin Waters and Oliver Wang and Joshua Ainslie and Jason Baldridge and Han Zhang and Garima Pruthi and Jakob Bauer and Feng Yang and Riham Mansour and Jason Gelman and Yang Xu and George Polovets and Ji Liu and Honglong Cai and Warren Chen and XiangHai Sheng and Emily Xue and Sherjil Ozair and Christof Angermueller and Xiaowei Li and Anoop Sinha and Weiren Wang and Julia Wiesinger and Emmanouil Koukoumidis and Yuan Tian and Anand Iyer and Madhu Gurumurthy and Mark Goldenson and Parashar Shah and MK Blake and Hongkun Yu and Anthony Urbanowicz and Jennimaria Palomaki and Chrisantha Fernando and Ken Durden and Harsh Mehta and Nikola Momchev and Elahe Rahimtoroghi and Maria Georgaki and Amit Raul and Sebastian Ruder and Morgan Redshaw and Jinhyuk Lee and Denny Zhou and Komal Jalan and Dinghua Li and Blake Hechtman and Parker Schuh and Milad Nasr and Kieran Milan and Vladimir Mikulik and Juliana Franco and Tim Green and Nam Nguyen and Joe Kelley and Aroma Mahendru and Andrea Hu and Joshua Howland and Ben Vargas and Jeffrey Hui and Kshitij Bansal and Vikram Rao and Rakesh Ghiya and Emma Wang and Ke Ye and Jean Michel Sarr and Melanie Moranski Preston and Madeleine Elish and Steve Li and Aakash Kaku and Jigar Gupta and Ice Pasupat and Da-Cheng Juan and Milan Someswar and Tejvi M. and Xinyun Chen and Aida Amini and Alex Fabrikant and Eric Chu and Xuanyi Dong and Amruta Muthal and Senaka Buthpitiya and Sarthak Jauhari and Nan Hua and Urvashi Khandelwal and Ayal Hitron and Jie Ren and Larissa Rinaldi and Shahar Drath and Avigail Dabush and Nan-Jiang Jiang and Harshal Godhia and Uli Sachs and Anthony Chen and Yicheng Fan and Hagai Taitelbaum and Hila Noga and Zhuyun Dai and James Wang and Chen Liang and Jenny Hamer and Chun-Sung Ferng and Chenel Elkind and Aviel Atias and Paulina Lee and Vít Listík and Mathias Carlen and Jan van de Kerkhof and Marcin Pikus and Krunoslav Zaher and Paul Müller and Sasha Zykova and Richard Stefanec and Vitaly Gatsko and Christoph Hirnschall and Ashwin Sethi and Xingyu Federico Xu and Chetan Ahuja and Beth Tsai and Anca Stefanoiu and Bo Feng and Keshav Dhandhania and Manish Katyal and Akshay Gupta and Atharva Parulekar and Divya Pitta and Jing Zhao and Vivaan Bhatia and Yashodha Bhavnani and Omar Alhadlaq and Xiaolin Li and Peter Danenberg and Dennis Tu and Alex Pine and Vera Filippova and Abhipso Ghosh and Ben Limonchik and Bhargava Urala and Chaitanya Krishna Lanka and Derik Clive and Yi Sun and Edward Li and Hao Wu and Kevin Hongtongsak and Ianna Li and Kalind Thakkar and Kuanysh Omarov and Kushal Majmundar and Michael Alverson and Michael Kucharski and Mohak Patel and Mudit Jain and Maksim Zabelin and Paolo Pelagatti and Rohan Kohli and Saurabh Kumar and Joseph Kim and Swetha Sankar and Vineet Shah and Lakshmi Ramachandruni and Xiangkai Zeng and Ben Bariach and Laura Weidinger and Tu Vu and Alek Andreev and Antoine He and Kevin Hui and Sheleem Kashem and Amar Subramanya and Sissie Hsiao and Demis Hassabis and Koray Kavukcuoglu and Adam Sadovsky and Quoc Le and Trevor Strohman and Yonghui Wu and Slav Petrov and Jeffrey Dean and Oriol Vinyals},
      year={2025},
      eprint={2312.11805},
      archivePrefix={arXiv},
      primaryClass={cs.CL},
      url={https://arxiv.org/abs/2312.11805}, 
}

@article{Yang_2013,
    author = "Yang, J. C. and others",
    editor = "Sakurai, Hiroyoshi and Miyatake, Hiroari and Xiao, Guoqing and Uesaka, Tomohiro and Huang, Wenxue",
    title = "{High Intensity heavy ion Accelerator Facility (HIAF) in China}",
    doi = "10.1016/j.nimb.2013.08.046",
    journal = "Nucl. Instrum. Meth. B",
    volume = "317",
    pages = "263--265",
    year = "2013"
}

@article{Schatz_2022,
   title={Horizons: nuclear astrophysics in the 2020s and beyond},
   volume={49},
   ISSN={1361-6471},
   url={http://dx.doi.org/10.1088/1361-6471/ac8890},
   DOI={10.1088/1361-6471/ac8890},
   number={11},
   journal={Journal of Physics G: Nuclear and Particle Physics},
   publisher={IOP Publishing},
   author={Schatz, H and Becerril Reyes, A D and Best, A and Brown, E F and Chatziioannou, K and Chipps, K A and Deibel, C M and Ezzeddine, R and Galloway, D K and Hansen, C J and Herwig, F and Ji, A P and Lugaro, M and Meisel, Z and Norman, D and Read, J S and Roberts, L F and Spyrou, A and Tews, I and Timmes, F X and Travaglio, C and Vassh, N and Abia, C and Adsley, P and Agarwal, S and Aliotta, M and Aoki, W and Arcones, A and Aryan, A and Bandyopadhyay, A and Banu, A and Bardayan, D W and Barnes, J and Bauswein, A and Beers, T C and Bishop, J and Boztepe, T and Côté, B and Caplan, M E and Champagne, A E and Clark, J A and Couder, M and Couture, A and de Mink, S E and Debnath, S and deBoer, R J and den Hartogh, J and Denissenkov, P and Dexheimer, V and Dillmann, I and Escher, J E and Famiano, M A and Farmer, R and Fisher, R and Fröhlich, C and Frebel, A and Fryer, C and Fuller, G and Ganguly, A K and Ghosh, S and Gibson, B K and Gorda, T and Gourgouliatos, K N and Graber, V and Gupta, M and Haxton, W C and Heger, A and Hix, W R and Ho, W C G and Holmbeck, E M and Hood, A A and Huth, S and Imbriani, G and Izzard, R G and Jain, R and Jayatissa, H and Johnston, Z and Kajino, T and Kankainen, A and Kiss, G G and Kwiatkowski, A and La Cognata, M and Laird, A M and Lamia, L and Landry, P and Laplace, E and Launey, K D and Leahy, D and Leckenby, G and Lennarz, A and Longfellow, B and Lovell, A E and Lynch, W G and Lyons, S M and Maeda, K and Masha, E and Matei, C and Merc, J and Messer, B and Montes, F and Mukherjee, A and Mumpower, M R and Neto, D and Nevins, B and Newton, W G and Nguyen, L Q and Nishikawa, K and Nishimura, N and Nunes, F M and O’Connor, E and O’Shea, B W and Ong, W-J and Pain, S D and Pajkos, M A and Pignatari, M and Pizzone, R G and Placco, V M and Plewa, T and Pritychenko, B and Psaltis, A and Puentes, D and Qian, Y-Z and Radice, D and Rapagnani, D and Rebeiro, B M and Reifarth, R and Richard, A L and Rijal, N and Roederer, I U and Rojo, J S and J S K and Saito, Y and Schwenk, A and Sergi, M L and Sidhu, R S and Simon, A and Sivarani, T and Skúladóttir, Á and Smith, M S and Spiridon, A and Sprouse, T M and Starrfield, S and Steiner, A W and Strieder, F and Sultana, I and Surman, R and Szücs, T and Tawfik, A and Thielemann, F and Trache, L and Trappitsch, R and Tsang, M B and Tumino, A and Upadhyayula, S and Valle Martínez, J O and Van der Swaelmen, M and Viscasillas Vázquez, C and Watts, A and Wehmeyer, B and Wiescher, M and Wrede, C and Yoon, J and Zegers, R G T and Zermane, M A and Zingale, M},
   year={2022},
   month=Nov, pages={110502} }

@article{Caballero_2016,
  title = {First Measurement of Several $\ensuremath{\beta}$-Delayed Neutron Emitting Isotopes Beyond $N=126$},
  author = {Caballero-Folch, R. and Domingo-Pardo, C. and Agramunt, J. and Algora, A. and Ameil, F. and Arcones, A. and Ayyad, Y. and Benlliure, J. and Borzov, I. N. and Bowry, M. and Calvi\~no, F. and Cano-Ott, D. and Cort\'es, G. and Davinson, T. and Dillmann, I. and Estrade, A. and Evdokimov, A. and Faestermann, T. and Farinon, F. and Galaviz, D. and Garc\'{\i}a, A. R. and Geissel, H. and Gelletly, W. and Gernh\"auser, R. and G\'omez-Hornillos, M. B. and Guerrero, C. and Heil, M. and Hinke, C. and Kn\"obel, R. and Kojouharov, I. and Kurcewicz, J. and Kurz, N. and Litvinov, Yu. A. and Maier, L. and Marganiec, J. and Marketin, T. and Marta, M. and Mart\'{\i}nez, T. and Mart\'{\i}nez-Pinedo, G. and Montes, F. and Mukha, I. and Napoli, D. R. and Nociforo, C. and Paradela, C. and Pietri, S. and Podoly\'ak, Zs. and Prochazka, A. and Rice, S. and Riego, A. and Rubio, B. and Schaffner, H. and Scheidenberger, Ch. and Smith, K. and Sokol, E. and Steiger, K. and Sun, B. and Ta\'{\i}n, J. L. and Takechi, M. and Testov, D. and Weick, H. and Wilson, E. and Winfield, J. S. and Wood, R. and Woods, P. and Yeremin, A.},
  journal = {Phys. Rev. Lett.},
  volume = {117},
  issue = {1},
  pages = {012501},
  numpages = {6},
  year = {2016},
  month = {Jun},
  publisher = {American Physical Society},
  doi = {10.1103/PhysRevLett.117.012501},
  url = {https://link.aps.org/doi/10.1103/PhysRevLett.117.012501}
}

@article{Phong_2022,
  title = {$\ensuremath{\beta}$-Delayed One and Two Neutron Emission Probabilities Southeast of $^{132}\mathrm{Sn}$ and the Odd-Even Systematics in $r$-Process Nuclide Abundances},
  author = {Phong, V. H. and Nishimura, S. and Lorusso, G. and Davinson, T. and Estrade, A. and Hall, O. and Kawano, T. and Liu, J. and Montes, F. and Nishimura, N. and Grzywacz, R. and Rykaczewski, K. P. and Agramunt, J. and Ahn, D. S. and Algora, A. and Allmond, J. M. and Baba, H. and Bae, S. and Brewer, N. T. and Bruno, C. G. and Caballero-Folch, R. and Calvi\~no, F. and Coleman-Smith, P. J. and Cortes, G. and Dillmann, I. and Domingo-Pardo, C. and Fijalkowska, A. and Fukuda, N. and Go, S. and Griffin, C. J. and Ha, J. and Harkness-Brennan, L. J. and Isobe, T. and Kahl, D. and Khiem, L. H. and Kiss, G. G. and Korgul, A. and Kubono, S. and Labiche, M. and Lazarus, I. and Liang, J. and Liu, Z. and Matsui, K. and Miernik, K. and Moon, B. and Morales, A. I. and Morrall, P. and Nepal, N. and Page, R. D. and Piersa-Si\l{}kowska, M. and Pucknell, V. F. E. and Rasco, B. C. and Rubio, B. and Sakurai, H. and Shimizu, Y. and Stracener, D. W. and Sumikama, T. and Suzuki, H. and Tain, J. L. and Takeda, H. and Tarife\~no-Saldivia, A. and Tolosa-Delgado, A. and Woli\ifmmode \acute{n}\else \'{n}\fi{}ska-Cichocka, M. and Woods, P. J. and Yokoyama, R.},
  journal = {Phys. Rev. Lett.},
  volume = {129},
  issue = {17},
  pages = {172701},
  numpages = {7},
  year = {2022},
  month = {Oct},
  publisher = {American Physical Society},
  doi = {10.1103/PhysRevLett.129.172701},
  url = {https://link.aps.org/doi/10.1103/PhysRevLett.129.172701}
}

@article{Lorusso_2015,
  title = {$\ensuremath{\beta}$-Decay Half-Lives of 110 Neutron-Rich Nuclei across the $N=82$ Shell Gap: Implications for the Mechanism and Universality of the Astrophysical $r$ Process},
  author = {Lorusso, G. and Nishimura, S. and Xu, Z. Y. and Jungclaus, A. and Shimizu, Y. and Simpson, G. S. and S\"oderstr\"om, P.-A. and Watanabe, H. and Browne, F. and Doornenbal, P. and Gey, G. and Jung, H. S. and Meyer, B. and Sumikama, T. and Taprogge, J. and Vajta, Zs. and Wu, J. and Baba, H. and Benzoni, G. and Chae, K. Y. and Crespi, F. C. L. and Fukuda, N. and Gernh\"auser, R. and Inabe, N. and Isobe, T. and Kajino, T. and Kameda, D. and Kim, G. D. and Kim, Y.-K. and Kojouharov, I. and Kondev, F. G. and Kubo, T. and Kurz, N. and Kwon, Y. K. and Lane, G. J. and Li, Z. and Montaner-Piz\'a, A. and Moschner, K. and Naqvi, F. and Niikura, M. and Nishibata, H. and Odahara, A. and Orlandi, R. and Patel, Z. and Podoly\'ak, Zs. and Sakurai, H. and Schaffner, H. and Schury, P. and Shibagaki, S. and Steiger, K. and Suzuki, H. and Takeda, H. and Wendt, A. and Yagi, A. and Yoshinaga, K.},
  journal = {Phys. Rev. Lett.},
  volume = {114},
  issue = {19},
  pages = {192501},
  numpages = {7},
  year = {2015},
  month = {May},
  publisher = {American Physical Society},
  doi = {10.1103/PhysRevLett.114.192501},
  url = {https://link.aps.org/doi/10.1103/PhysRevLett.114.192501}
}

@article{Borge_2017,
author = {Borge, Maria and Jonson, B.},
year = {2017},
month = {03},
pages = {044011},
title = {ISOLDE past, present and future},
volume = {44},
journal = {Journal of Physics G: Nuclear and Particle Physics},
doi = {10.1088/1361-6471/aa5f03}
}

@article{GARNSWORTHY_2019,
title = {The GRIFFIN facility for Decay-Spectroscopy studies at TRIUMF-ISAC},
journal = {Nuclear Instruments and Methods in Physics Research Section A: Accelerators, Spectrometers, Detectors and Associated Equipment},
volume = {918},
pages = {9-29},
year = {2019},
issn = {0168-9002},
doi = {https://doi.org/10.1016/j.nima.2018.11.115},
url = {https://www.sciencedirect.com/science/article/pii/S0168900218317662},
author = {A.B. Garnsworthy and C.E. Svensson and M. Bowry and R. Dunlop and A.D. MacLean and B. Olaizola and J.K. Smith and F.A. Ali and C. Andreoiu and J.E. Ash and W.H. Ashfield and G.C. Ball and T. Ballast and C. Bartlett and Z. Beadle and P.C. Bender and N. Bernier and S.S. Bhattacharjee and H. Bidaman and V. Bildstein and D. Bishop and P. Boubel and R. Braid and D. Brennan and T. Bruhn and C. Burbadge and A. Cheeseman and A. Chester and R. Churchman and S. Ciccone and R. Caballero-Folch and D.S. Cross and S. Cruz and B. Davids and A. Diaz Varela and I. Dillmann and M.R. Dunlop and L.J. Evitts and F.H. Garcia and P.E. Garrett and S. Georges and S. Gillespie and R. Gudapati and G. Hackman and B. Hadinia and S. Hallam and J. Henderson and S.V. Ilyushkin and B. Jigmeddorj and A.I. Kilic and D. Kisliuk and R. Kokke and K. Kuhn and R. Krücken and M. Kuwabara and A.T. Laffoley and R. Lafleur and K.G. Leach and J.R. Leslie and Y. Linn and C. Lim and E. MacConnachie and A.R. Mathews and E. McGee and J. Measures and D. Miller and W.J. Mills and W. Moore and D. Morris and L.N. Morrison and M. Moukaddam and C.R. Natzke and K. Ortner and E. Padilla-Rodal and O. Paetkau and J. Park and H.P. Patel and C.J. Pearson and E. Peters and E.E. Peters and J.L. Pore and A.J. Radich and M.M. Rajabali and E.T. Rand and K. Raymond and U. Rizwan and P. Ruotsalainen and Y. Saito and F. Sarazin and B. Shaw and J. Smallcombe and D. Southall and K. Starosta and M. Ticu and E. Timakova and J. Turko and R. Umashankar and C. Unsworth and Z.M. Wang and K. Whitmore and S. Wong and S.W. Yates and E.F. Zganjar and T. Zidar}
}

@ARTICLE{Tanaka_2013,
       author = {{Tanaka}, Masaomi and {Hotokezaka}, Kenta},
        title = "{Radiative Transfer Simulations of Neutron Star Merger Ejecta}",
      journal = {\apj},
         year = 2013,
        month = oct,
       volume = {775},
       number = {2},
          eid = {113},
        pages = {113},
          doi = {10.1088/0004-637X/775/2/113},
archivePrefix = {arXiv},
       eprint = {1306.3742},
 primaryClass = {astro-ph.HE},
       adsurl = {https://ui.adsabs.harvard.edu/abs/2013ApJ...775..113T}
}

\end{document}